\documentclass[12pt]{article}
\usepackage{amssymb}
\newif\ifusepdf
\usepdffalse  
\ifusepdf
\newif\ifpdf
\ifx\pdfoutput\undefined
   \pdffalse
   \usepackage{cite}
 \else
   \pdfoutput=1
   \pdftrue
  \usepackage[pdftex]{hyperref}
   \providecommand{\href}[2]{#2}
\newcommand\email[1]{{\tt\href{mailto:#1}{\textbf{\texttt{#1}}}}}
\fi
\else
\newcommand\email[1]{\textbf{\texttt{#1}}}
\newcommand\href[2]{#2}
\fi

\csname @addtoreset\endcsname{equation}{section}

\newcommand{\be}{\begin{equation}}

\newcommand{\ee}{\end{equation}}

\newcommand{\bea}{\begin{eqnarray}}
\newcommand{\eea}{\end{eqnarray}}

 \def\slash#1{\setbox0=\hbox{$#1$}#1\hskip-\wd0\dimen0=5pt\advance
       \dimen0 by-\ht0\advance\dimen0 by\dp0\lower0.5\dimen0\hbox
         to\wd0{\hss\sl/\/\hss}}

\renewcommand{\a}{\alpha}
\renewcommand{\b}{\beta}

\newcommand{\e}{\epsilon}
\newcommand{\m}{\mu}
\newcommand{\n}{\nu}

\newcommand{\bi}{\begin{enumerate}}
\newcommand{\ei}{\end{enumerate}}

      \newcommand{\s}{\sigma}

\def\6{\partial}
\def\7{\tilde}
\def\8{\hat}

\def\={{\;=\;}}\def\+{{\;+\;}}

\renewcommand{\varepsilon }{\epsilon }
\newcommand{\dslash}{\not\hspace{-.05cm}\partial}
\newcommand{\Dslash}{\not\hspace{-.1cm}D}
\begin{document}
\begin{titlepage}
\begin{flushright}
arXiv:yymm.nnnn [hep-th]
\end{flushright}
\vspace{.5cm}
\begin{center}
\baselineskip=16pt {\LARGE  \bf{Field Representations of \\ Vector Supersymmetry}}
\vskip 5mm

{\large Roberto Casalbuoni$^1$, Federico Elmetti$^2$,
 \\[2mm]
Simon Knapen$^{2,3}$ and Laura Tamassia$^2$}
\vskip 5mm

{\small {\it $^1$ Department of Physics, University of Florence,\\
         INFN-Florence and Galileo Galilei Institute,
          Florence, Italy} \\
          \email{casalbuoni@fi.infn.it}\\\vspace{4mm}
$^2$ {\it Instituut voor Theoretische Fysica, Katholieke Universiteit Leuven,\\
Celestijnenlaan 200D B-3001 Leuven, Belgium} \\
\email{federico.elmetti@alice.it},
\email{laura.tamassia@fys.kuleuven.be}, \\\vspace{4mm}
$^3$ {\it Department of Physics and Astronomy, Rutgers University, \\
Piscataway, NJ 08855-0849, USA} \\
\email{knapen@physics.rutgers.edu}
\\\vspace{1mm}}

\end{center}
\vskip 3mm
\begin{center}
{\bf Abstract}
\end{center}
{\small We study some field representations of vector supersymmetry with superspin $Y=0$ and $Y=1/2$ and nonvanishing central charges.
For $Y=0$, we present two multiplets composed of four spinor fields, two even and two odd, and we provide a free action for them.
The main differences between these two multiplets are the way the central charge operators act and the compatibility with the Majorana reality condition on the spinors. One of the two is related to a previously studied spinning particle model. 
For $Y=1/2$, we present a multiplet composed of one even scalar, one odd vector and one even selfdual two-form,
which is a truncation of a known representation of the tensor supersymmetry algebra in Euclidean spacetime. 
We discuss its rotation to Minkowski spacetime and provide a set of dynamical equations for it, which are however not derived from a Lagrangian. We develop a superspace formalism for vector supersymmetry with central charges and we derive our multiplets by superspace techniques.
Finally, we discuss some representations with vanishing central charges.

}
\end{titlepage}
\addtocounter{page}{1}
 \tableofcontents{}
\newpage

\section{Introduction\label{sec:0}}

In this paper we  study some field representations of the vector supersymmetry algebra, a graded extension of the
Poincar{\'e} algebra in four dimensions. The extension is realized by adding to the Poincar\'e algebra two fermionic operators,
an odd Lorentz vector and an odd Lorentz scalar. Furthermore,  two central charges
are allowed. The anticommutator between vector and scalar odd generators
gives the four-momentum vector, from which the name vector supersymmetry (VSUSY).

To our knowledge, this algebra was first introduced in \cite{Barducci:1976qu} in 1976, with
the purpose of obtaining a pseudoclassical description of the Dirac
equation. Its general algebraic properties have been studied in \cite{Casalbuoni:2008ez} (see also \cite{Casalbuoni:2009en}).
The VSUSY algebra, or, better, an extension of it, arises in the context
 of topological
field theories. In fact, an Euclidean version of VSUSY appears as a
subalgebra of the symmetry algebra underlying topological $\mathcal{N}=2$
Yang-Mills theories. Supersymmetry with odd vector generators was studied
after Witten \cite{Witten:1988ze}, who, in 1988, introduced topological
$\mathcal{N}=2$ Yang-Mills theories by performing a topological twist.
After this twist, the fermionic generators become a vector, a scalar and
an anti-selfdual tensor \cite{Alvarez:1994ii,Kato:2005fj}. After
truncation of the anti-selfdual sector, the twisted algebra coincides
with the Euclidean VSUSY algebra, in the special case when the two
central charges of VSUSY are identified. Twisted topological algebras
have proven to be useful in the study of renormalization properties of
topological field theories \cite{Birmingham:1988bx,Delduc:1989ft}.
Moreover, a superspace formalism has been developed for these topological
theories, see for example
\cite{Alvarez:1994ii,Kato:2005fj,Baulieu:2008at} and references therein.

The main difference between vector and ordinary supersymmetry is that the
odd generators of VSUSY have integer spin and so they do not satisfy the usual spin-statistics relation. This implies that in any representation of VSUSY some of the component fields, counting for half of the degrees of freedom of the multiplet, necessarily violate the usual spin-statistics relation and in a field theory setting should be identified with ghosts. VSUSY then unifies physical fields with ghosts and not fields of integer spin with fields of half-integer spin. In particular, a VSUSY multiplet always contains either only fields of integer spin or only fields of half-integer spin.
Therefore, while VSUSY's algebraic structure is very similar to the one of ordinary supersymmetry, it has completely different implications.  Ghost fields are not observed but are nonetheless a very important technical tool in field and string theories. For this reason, VSUSY representations and dynamical models with underlying VSUSY are worth exploring. In any case, it
is interesting to compare this alternative to ordinary
supersymmetry to understand what the essential ingredients in
supersymmetry are.

In \cite{Casalbuoni:2008ez}, it has been shown that the irreducible representations of VSUSY can be classified according to the value assumed by the superspin Casimir operator, in complete analogy to the case of standard supersymmetry. An irreducible multiplet with a given value of the superspin $Y$ contains components of Lorentz spin $s=|Y\pm \frac 12|$ for $Y\not =0$, whereas for $Y=0$ the components have Lorentz spin $\frac 1 2$.

In this paper we explicitly construct field representations corresponding to the two lowest superspin values, $Y=0$ and $Y= \frac 12$. 
In the case $Y=0$, we find two off-shell multiplets both composed of four Dirac spinors, two even and two odd. The first one is compatible with Dirac-type equations of motion and is realized in the spinning particle model in \cite{Casalbuoni:2008iy}. The second one differs from the first because of the nontrivial action of the central charge operator on the fields. Moreover, it is not compatible with Dirac equations of motion, but only with Klein-Gordon ones.  
In the case $Y=\frac 12$, in Euclidean space, we construct a multiplet with a scalar, a vector and a selfdual two-form. The scalar and the two-form are even and the vector is odd, or the other way around. We show that this is a truncation of a representation of the topologically twisted $\mathcal {N}=2$ theory given in \cite{Kato:2005gb} (see also \cite{Kato:2008dw}).  Since in this paper we are mainly interested in representations in Minkowski spacetime, we discuss the existence of a similar multiplet in Minkowskian signature. We find that in Minkowski spacetime the number of degrees of freedom must be doubled, since the fields must be necessarily complex. The resulting Minkowskian multiplet features a complex scalar, a complex vector and a selfdual and an antiselfdual two-forms related to each other by complex conjugation. This is of course equivalent to having two real scalars, two real vectors and one real two-form.
For all multiplets we give the invariant free dynamical equations, which for $Y=0$ multiplets can also be derived from an action.

Furthermore, we develop a superspace setup for VSUSY and we rederive our results in components from superfields. To do that, we adapt the superspace with central charge first introduced by Sohnius in \cite{Sohnius:1978fw} to VSUSY. This kind of extended superspace allows us to also derive multiplets where the central charge operators act nontrivially on the fields.
Concerning the derivation of actions from superspace, we only take a first step by choosing to work with a fixed value of the central charge. In this way we derive the action for one of our $Y=0$ multiplets.

Finally, we consider the case of vanishing central charges and we find some examples of field representations by superspace techniques.

In the most interesting case of nonvanishing central charges, we limit our study to free fields. We leave interacting theories for future work.
In the case with vanishing central charge we find an interacting action for one of our multiplets. However, this action has the unusual property of being odd. The need for this kind of actions for some supersymmetric models in the case of ordinary supersymmetry has already been pointed out in \cite{Soroka:1995et}, \cite{Soroka:2001jg}. However, the quantization of these models remains to our knowledge problematic and goes in any case beyond the scope of this paper.

Part of the results presented in this paper are also discussed in the master thesis of Simon Knapen \cite{thesis}.
\vskip 12pt

\noindent The paper is organized as follows.\\
In Section 2 we state our conventions and we review the results obtained in \cite{Casalbuoni:2008ez} with particular emphasis on the classification of irreducible representations of VSUSY in the case of nonvanishing central charges.\\
In Section 3 we present two multiplets with $Y=0$. Furthermore, we introduce VSUSY superspace with central charges and we derive both multiplets by superspace techniques.\\
In Section 4 we present a $Y=\frac 12$ Euclidean on-shell multiplet, we discuss its rotation to Minskowski spacetime and its superspace origin.\\
In Section 5 we initiate the study of the construction of VSUSY invariant actions by superspace techniques by considering the case of the spinning particle $Y=0$ multiplet.\\
In Section 6 we study some representations of VSUSY with vanishing central charges.\\
In Section 7 we give our conclusions and outlook.\\
In Appendix A we give more details on the relation between one of our multiplets with $Y=0$ and the spinning particle model of \cite{Casalbuoni:2008iy}.\\
In Appendix B we explicitly show how to solve the first superspace constraint equation encountered in Section 3.\\
In Appendix C we discuss the relation between the $Y=1/2$ VSUSY multiplet of Section 4 and a tensor supersymmetry multiplet found in \cite{Kato:2005gb} in the context of ${\cal N}=2$ twisted topological models.
\section{VSUSY algebra and Casimir operators}
\subsection{Conventions}
Vector supersymmetry (VSUSY) algebra is a graded extension of the Poincar{\'e} algebra in four dimensions. The extension is realized by adding to the Poincar\'e algebra two fermionic operators, an odd Lorentz vector and an odd Lorentz scalar. Furthermore,  two central charges
are allowed. We work with Minkowski metric unless stated otherwise. The VSUSY commutation relations used are:
\begin{eqnarray}
&&[M_{\mu\nu},M_{\rho\sigma}]=-i\eta_{\nu\rho}M_{\mu\sigma}-
i\eta_{\mu\sigma}M_{\nu\rho}+i\eta_{\nu\sigma}M_{\mu\rho}+
i\eta_{\mu\rho}M_{\nu\sigma}\,;
\nonumber\\
&&[M_{\mu\nu},P_\rho]=i\eta_{\mu\rho}P_\nu-i\eta_{\nu\rho}P_\mu\,;\qquad
[M_{\mu\nu},Q_\rho]=i\eta_{\mu\rho}Q_\nu-i\eta_{\nu\rho}Q_\mu\,;
\nonumber\\
&&\{Q_{\mu},Q_{\nu}\}=Z \eta_{\mu\nu}\,;\quad \quad \{Q_{5},Q_{5}\}=\tilde{Z}\,;\quad \quad \{Q_{\mu},Q_{5}\}=-P_{\mu}\,.
\label{VSUSYalgebra}
\end{eqnarray}
In \cite{Casalbuoni:2008ez} it was shown that the VSUSY algebra can be derived by contraction from the simple orthosymplectic algebra $\rm OSp(3,2|2)$. The name we have chosen for the scalar odd generator, $Q_{5}$, is a reminder of its five-dimensional origin.
The algebra will be realized in the
 form (\ref{VSUSYalgebra}) in terms of differential operators acting on superfields. As usual, it will be realized with opposite signs in the RHS on component fields.\\
We follow the conventions:
\begin{eqnarray}
&\eta_{\mathcal{\mu\nu}} = \rm diag\{-1,+1,+1,+1\}\,;\quad \quad \e^{0123}=-\e_{0123}=1\,; \nonumber\\
&P_{\mu}= i\partial_{\mu}\,;\quad \quad \Box \equiv \partial^{\mu} \partial_{\mu}= \dslash \dslash = -P^{2}\,;\quad\quad\{\gamma_{\mu},\gamma_{\nu}\}=2\eta_{\mathcal{\mu\nu}}\,;\nonumber\\
&\gamma_{0}^{\dag}=-\gamma_{0}\,; \quad \quad \gamma_{i}^{\dag}=\gamma_{i}\,; \quad \quad \gamma_{0}^{-1}\gamma_{\mu}^{\dag}\gamma_{0}=-\gamma_{\mu}\,; \quad\quad \gamma_{5}\equiv \frac{i}{4!}\epsilon_{\mu\nu\rho\sigma}\gamma^{\mu}\gamma^{\nu}\gamma^{\rho}\gamma^{\sigma}\,.\cr
&
\label{conv}
\end{eqnarray}
Therefore we have the mass shell condition $P^2=-m^2$ and the Dirac equation is of the form $(\dslash +m)\psi=0$.\\
In this paper we discuss multiplets including both physical fields and ghosts. In general, to avoid confusion, we will denote fermionic fields by a tilde .\\
For our conventions concerning spinor fields, discussed next, we follow \cite{VanProeyen:1999ni}.\\
For the complex conjugation of a bilinear in two fermionic fields $\tilde A$ and $\tilde B$, we adopt the convention of exchanging the position of the fields, i.e.
\begin{equation}
 \left(\tilde{A}\tilde{B}\right)^\dagger=\tilde{B}^\dagger \tilde{A}^\dagger\,.
\label{fermionbil}
\end{equation}
We define the infinitesimal VSUSY transformations of the fields as
\begin{equation}
\delta_{Q_{5}} \psi=i\e_5 Q_{5}\psi~~~{\rm and}~~~ \delta_{Q_{\m}}\psi=i\e^{\m} Q_{\m} \psi.
\label{fulltrans}
\end{equation}
with  $\e_5$ and $\epsilon_\m $ real odd parameters. For the central charge operators, we define
\begin{equation}
\delta_{Z} \psi=i\e_Z Z\psi~~~{\rm and}~~~ \delta_{\tilde Z}\psi=i\e_{\tilde Z} \tilde{Z} \psi.
\label{fulltransZ}
\end{equation}
where the even parameters $\e_Z$ and $\e_{\tilde Z}$ are also taken to be real.
For Dirac conjugate fields and their transformations, we have the following conventions
\begin{eqnarray}
&\bar\psi= i\psi^\dagger \gamma_0\,, \nonumber\\
&\delta_{Q} \psi^\dagger=(\delta_{Q} \psi)^\dagger=-i (Q\psi)^\dagger \e^\dagger\,.
\label{daggerconv}
\end{eqnarray}
We use the following definition of the Majorana condition for spinors:
\begin{equation}
 i\psi^{\dag}\gamma_{0}=\psi^{T}\mathcal{C},
\label{majorana}
\end{equation}
where the charge conjugation matrix $\mathcal{C}$ satisfies
\begin{equation}
\mathcal{C}^{T}=-\mathcal{C}\,; \quad \quad \mathcal{C}^{-1}\gamma_{\mu}^{T}\mathcal{C}=-\gamma_{\mu}.
\end{equation}
As a result, for anti-commuting Majorana spinors we have the following identities:
\begin{equation}
\overline{\tilde{\chi}}\tilde{\xi}=\overline{\tilde{\xi}}\tilde{\chi}\,; \quad \quad \quad \quad \overline{\tilde{\chi}}\gamma_{\mu}\tilde{\xi}=-\overline{\tilde{\xi}}\gamma_{\mu}\tilde{\chi}\,; \quad \quad \quad \quad\int \overline{\tilde{\chi}}\dslash\tilde{\xi}=\int \overline{\tilde{\xi}}\dslash\tilde{\chi},
\end{equation}
while for commuting Majorana spinors we have:
\begin{equation}
\overline{\psi}\lambda=-\overline{\lambda}\psi\,; \quad \quad \quad \quad \overline{\psi}\gamma_{\mu}\lambda=\overline{\lambda}\gamma_{\mu}\psi\,;   \quad \quad \quad \quad \int \overline{\psi}\dslash\lambda=-\int \overline{\lambda}\dslash\psi.
\end{equation}
In practice, it is useful to rephrase (\ref{majorana}) in terms of the C-operation defined as follows on a spinor $\psi$
\begin{equation}
\psi^C=iB^{-1}\psi^* ~,~ {\rm where}~~~ B=-\mathcal{C} \gamma_0\,.
\label{Cop}
\end{equation}
The Majorana condition is then simply rewritten in the form
\begin{equation}
 \psi=\psi^C \,.
\end{equation}
In this case we say that the spinor $\psi$ is real.
For a general matrix $M$ in spinor space the C-operation is defined as follows:
\begin{equation}
M^C= B^{-1}M^* B\,.
\end{equation}
The gamma matrices behave as real matrices 
\begin{equation}
 (\gamma_\m)^C=\gamma_\m
\end{equation}
and the matrix $\gamma_5$ as purely imaginary
\begin{equation}
 (\gamma_5)^C=-\gamma_5\,.
\end{equation}
In the following, we will be interested in checking whether our spinorial VSUSY multiplets are compatible with a Majorana reality condition on the spinors.
A multiplet is compatible with the Majorana condition when for a real spinor $\psi$ the variation $\delta\psi$ is also real. Since we are dealing with both fermionic and bosonic spinors in this paper, the Majorana condition must be investigated separately in the two cases.
If $\psi$ is real and {\em odd} (physical), then requiring compatibility with the Majorana condition is equivalent to asking that $Q_\m \psi$ and $Q_5 \psi$ are purely imaginary.
If $\psi$ is real and {\em even} (ghost), then requiring compatibility with the Majorana condition is equivalent to asking that $Q_\m \psi$ and $Q_5 \psi$ are real. 
From the algebra (\ref{VSUSYalgebra}), $\{Q_\m,Q_\n\}=\eta_{\m\n} Z$, one can easily derive that $Z\psi$ has to be purely imaginary when $\psi$ is real, both in the even and odd cases. An analogue conclusion can be drawn for $\tilde Z$. Therefore the factor of $i$ present in the definitions (\ref{fulltransZ}) is needed for consistency of the reality conditions on the spinors with the algebra.

\subsection{Equivalence classes of irreducible representations with different values of the central charges}
As already noted in \cite{Casalbuoni:2008ez}, for irreducible representations where, in a suitable basis, the central charges can be treated as numbers, it is possible to make a rescaling such that in general only the value of one central charge and the relative sign between the two is relevant.
In fact, by implementing into the algebra (\ref{VSUSYalgebra}) the rescalings
\begin{equation}
 Q_\m \rightarrow \frac{1}{\a} Q_\m\,, \quad\quad Q_5 \rightarrow \a Q_5\,,
\end{equation}
one obtains
\begin{equation}
\{Q_\m,Q_\n\}\rightarrow \frac{1}{\a^2} \eta_{\m\n} Z, \quad\quad \{Q_5,Q_5\}=\a^2 \tilde Z\,,
\end{equation}
while the other (anti)commutation relations remain unchanged.
By choosing $\a^2=\sqrt{\frac{\vert Z\vert}{\vert \tilde Z\vert}}$, one sees that only the absolute value of one of the two central charges and their relative signs are relevant.
A similar rescaling can be performed when one of the central charges is zero and the other is not, to fix the value of the nonzero central charge to $\pm 1$.\\
To summarize, one gets the different equivalence classes of irreducible representations given in Table \ref{tableeqclass}.
\begin{table}[h]
\begin{center}
\begin{tabular}{|c|c|}
\hline
Central charge values & Equivalent with \\
\hline
&\\
$Z\neq 0$, $\tilde Z\neq 0$ and ${\rm sign}(Z)={\rm sign}(\tilde Z)$ &$Z=\tilde Z\neq 0$\\
$Z\neq 0$, $\tilde Z\neq 0$ and ${\rm sign}(Z)=-{\rm sign}(\tilde Z)$ &$Z=-\tilde Z\neq 0$\\
$Z=0$, $\tilde Z\neq 0$ & $Z=0$ and $\tilde Z=\pm 1$ \\
$Z\neq0$, $\tilde Z= 0$ & $Z=\pm 1$ and $\tilde Z=0$ \\
$Z=\tilde Z= 0$ & $Z=\tilde Z=0$ \\
&\\
\hline
\end{tabular}
\caption{Equivalence classes of irreducible representations according to the values of the central charges\label{tableeqclass}}
\end{center}
\end{table}
In some field representations we discuss in this paper, the two central charges of the VSUSY algebra are identified, for simplicity. One should then remember that this assumption is in fact a restriction only for reducible representations characterized by a nontrivial action of the central charge operators.
\subsection{Casimir operators}
In \cite{Casalbuoni:2008ez}, the Casimir operators of the VSUSY algebra (\ref{VSUSYalgebra}) have been derived.\\
Besides the square of the momentum $P^2$, another spin-related Casimir has been found, $W^{2}$, which is the square of the spin vector $W^\m$:
\begin{equation}
W^{\mu} = \frac{1}{2}\epsilon^{\mu\nu\rho\sigma}\left(i\,Z M_{\rho\sigma}-Q_{\rho}Q_{\sigma}\right).
\end{equation}
This is the VSUSY analogue of the superspin operator of ordinary supersymmetry. The structure of the superspin Casimir allows us to derive the Lorentz spin content of an irreducible VSUSY multiplet.\\
If one denotes the superspin eigenvalues by $W^{2}=m^2Y(Y+1)$, it turns out that an irreducible VSUSY multiplet of superspin $Y$ always contain fields of Lorentz spin $s=\vert Y\pm \frac{1}{2}\vert$. Therefore, for $Y>0$ the multiplets contain components of spin $(s,s+1)$, while in the degenerate case $Y=0$ the components have spin $\frac{1}{2}$. In this paper we will explicitly construct multiplets with $Y=0$ and $Y=\frac{1}{2}$.\\
In \cite{Casalbuoni:2008ez}, it was also observed that VSUSY, due to its structure with scalar and vectorial odd generators, could have an odd Casimir operator as well. In fact, it was shown that, for representations satisfying the following BPS-like condition relating the mass and the values of the central charges
\begin{equation}
P^{2}=Z\tilde Z\,,
\label{pzz}
\end{equation}
the odd operator 
\begin{equation}
\mathcal{Q}=Q^\m P_\m + Q_{5} Z,
\label{oddcas}
\end{equation}
(anti)commutes with all other operators in the algebra. Therefore, $\mathcal{Q}$ is technically not a Casimir operator, it should be called surface invariant. However, we still choose to call it `odd Casimir' in this paper.

\section{$Y=0$ multiplets}
According to the discussion in the previous section, the irreducible VSUSY multiplets with superspin $Y=0$ are doublets of spin $\frac{1}{2}$ fields. These doublets are realized in a field theory setting by on-shell fields. Off-shell multiplet have in general more fields. In the following we will present two inequivalent off-shell multiplets, both featuring four spinors, two even and two odd. For one of the multiplets we will also show a reduction, featuring only two spinors, one even and one odd, which closes only on shell. Furthermore, we will discuss whether these spinorial multiplets are compatible with a reality (Majorana) condition and we will give a VSUSY invariant free action for them.
Finally, we will provide a superspace setup for VSUSY and we will derive our multiplets by superspace techniques.
\subsection{$Y=0$ multiplet from the spinning particle}

Consider the four Dirac spinor fields $\psi_1$,  $\tilde\psi_1$, $\psi_2$, $\tilde\psi_2$, two even and two odd. As mentioned earlier, we denote fermionic fields by a tilde.
The VSUSY charges act on the fields as given in Table \ref{tablespinning}.
\begin{table}[h]
\begin{center}
\begin{tabular}{|c|c|c|}
    \hline
      & $Q_{5}$ & $Q_{\mu}$ \\
     \hline
      & & \\
    $\psi_1$ & $-\sqrt{\frac{a}{2}} \tilde\psi_2$ & $\frac{1}{\sqrt{2a}}\left[\left(-m\gamma_\m +  \partial_\m\right)\gamma_5\tilde\psi_1- i\partial_\m \tilde\psi_2\right]$\\
    $\tilde\psi_1$ & $-\sqrt{\frac{a}{2}} \psi_2$ & $\frac{1}{\sqrt{2a}}\left[\left(-m\gamma_\m +  \partial_\m\right)\gamma_5\psi_1- i\partial_\m \psi_2\right]$  \\
    $\psi_2$ & $-\sqrt{\frac{a}{2}} \tilde\psi_1$ & $\frac{1}{\sqrt{2a}}\left[- i\partial_\m \tilde\psi_1-\gamma_5\left(m\gamma_\m +  \partial_\m\right)\tilde\psi_2\right]$\\
    $\tilde\psi_2$ & $-\sqrt{\frac{a}{2}} \psi_1$ & $\frac{1}{\sqrt{2a}}\left[- i\partial_\m \psi_1-\gamma_5\left(m\gamma_\m +  \partial_\m\right)\psi_2\right]$  \\
     & & \\
    \hline
    \end{tabular}
\caption{Action of the VSUSY charges on the fields (four-spinor multiplet with fixed value of the central charges). \label{tablespinning}}
\end{center}
\end{table}
The central charges  $Z$ and $\tilde Z$ are represented diagonally on this multiplet, with values $Z=\frac{m^2}{a}$ and $\tilde Z=-a$.
The following set of Dirac-type equations of motion is invariant under VSUSY:
\begin{eqnarray}
 &\left(\dslash +m\right)\psi_1=0\,;~~~~~~~ \left(\dslash +m\right)\tilde\psi_1=0\,;\cr
&\left(\dslash +m\right)\psi_2=0\,;~~~~~~~ \left(\dslash +m\right)\tilde\psi_2=0\,.
\label{Diraceqs}
\end{eqnarray}
Now that the parameter $m$ in the representation has been identified with the mass of the spinors,
we see that the relation between the values of the two central charges $Z\tilde{Z}=-m^2$ is the one allowing for the presence of the odd Casimir on the mass shell (\ref{pzz}).
Indeed, one can explicitly check that the odd Casimir (\ref{oddcas}) is zero for this representation when $P^2=-m^2$, as expected in the absence of a natural odd constant in the model.\\
Following the conventions (\ref{daggerconv}), one can compute the transformations of the barred fields and the result is given in Table \ref{tablespinningbar}.
\begin{table}[h]
\begin{center}
\begin{tabular}{|c|c|c|}
    \hline
      & $\delta_{Q_{5}}$ & $\delta_{Q_\m}$ \\
     \hline
      & & \\
    $\bar\psi_1$ & $i\sqrt{\frac{a}{2}}\bar{\tilde\psi_2} \epsilon $ & $\frac{1}{\sqrt{2a}}\left[i\bar{\tilde\psi_1}\gamma_5\left(m\gamma_\m +  \overleftarrow\partial_\m\right)+ \bar{\tilde\psi_2}\overleftarrow\partial_\m \right]\e^\m$\\
    $\bar{\tilde\psi_1}$ & $i\sqrt{\frac{a}{2}} \bar\psi_2 \e $ & $\frac{1}{\sqrt{2a}}\left[i\bar\psi_1\gamma_5\left(m\gamma_\m + \overleftarrow\partial_\m\right)+ \bar\psi_2\overleftarrow\partial_\m\right] \e^\m$ \\
    $\bar\psi_2$ & $i\sqrt{\frac{a}{2}} \bar{\tilde\psi_1}\e$ & $\frac{1}{\sqrt{2a}}\left[ \bar{\tilde\psi_1}\overleftarrow\partial_\m +i\bar{\tilde\psi_2}\left(m\gamma_\m - \overleftarrow\partial_\m\right)\gamma_5\right] \e^\m$\\
    $\bar{\tilde\psi_2}$ & $i\sqrt{\frac{a}{2}}\bar\psi_1 \e$ & $\frac{1}{\sqrt{2a}}\left[ \bar\psi_1\overleftarrow\partial_\m +i\bar\psi_2\left(m\gamma_\m -  \overleftarrow\partial_\m\right)\gamma_5\right] \e^\m$  \\
     & & \\
    \hline
    \end{tabular}
\caption{VSUSY transformations of the Dirac-conjugate spinor fields (four-spinor multiplet with fixed value of the central charges).\label{tablespinningbar}}
\end{center}
\end{table}

\noindent One can check that the following Dirac-type action is VSUSY invariant
\begin{equation}
S=\int d^4x \left[-\bar\psi_2  \left(\dslash + m\right)\psi_1 + \bar{\tilde\psi_2}\left(\dslash +m\right)\tilde\psi_1 - \bar\psi_1\left(\dslash +m\right)\psi_2 +  \bar{\tilde\psi_1}\left(\dslash +m\right)\tilde\psi_2 \right].
\label{actionstandard}
\end{equation}
In \cite{Casalbuoni:2008iy}, a VSUSY invariant action was constructed for the massive spinning particle with the method of nonlinear realizations.
The representation found there can further be rewritten, as shown in Appendix A, in terms of four Dirac spinors $\psi_1$,  $\tilde\psi_1$, $\psi_2$, $\tilde\psi_2$, two even and two odd. One can easily show that that representation is the same as the one given in Table \ref{tablespinning}.
The action of the VSUSY charges $Q_{5}$ and $Q_\m$ on the spinors of \cite{Casalbuoni:2008iy} is given in Table \ref{tablespinorsold}.
\begin{table}[h]
\begin{center}
\begin{tabular}{|c|c|c|}
    \hline
      & $Q_{5}$ & $Q_{\mu}$ \\
     \hline
      & & \\
    $\psi_1$ & $-\sqrt{\frac{a}{2}}\gamma_5 \tilde\psi_2$ & $\frac{1}{\sqrt{2a}}\gamma_5\left[\left(-m\gamma_\m +  \partial_\m\right)\tilde\psi_1- i\partial_\m \tilde\psi_2\right]$\\
    $\tilde\psi_1$ & $-\sqrt{\frac{a}{2}}\gamma_5 \psi_2$ & $\frac{1}{\sqrt{2a}}\gamma_5\left[\left(-m\gamma_\m +  \partial_\m\right)\psi_1- i\partial_\m \psi_2\right]$  \\
    $\psi_2$ & $-\sqrt{\frac{a}{2}}\gamma_5 \tilde\psi_1$ & $\frac{1}{\sqrt{2a}}\gamma_5\left[- i\partial_\m \tilde\psi_1-\left(m\gamma_\m +  \partial_\m\right)\tilde\psi_2\right]$\\
    $\tilde\psi_2$ & $-\sqrt{\frac{a}{2}}\gamma_5 \psi_1$ & $\frac{1}{\sqrt{2a}}\gamma_5\left[- i\partial_\m \psi_1-\left(m\gamma_\m +  \partial_\m\right)\psi_2\right]$  \\
     & & \\
    \hline
    \end{tabular}
\end{center}
\caption{ Action of the VSUSY charges on the four-spinor multiplet in \cite{Casalbuoni:2008iy}. \label{tablespinorsold}}
\end{table}

\noindent To see that the representation in Table \ref{tablespinorsold} is in fact the same as the one given in Table \ref{tablespinning}, it is enough to implement the following rescaling
\begin{equation}
\psi_{1} \rightarrow \gamma_{5}\psi_{1}\,, \quad \quad \quad \quad \tilde\psi_{1} \rightarrow \gamma_{5}\tilde\psi_{1}.
\label{rescal}
\end{equation}
A natural question is whether the number of degrees of freedom in this multiplet could be reduced by half by imposing a reality (Majorana) condition on the spinors. The considerations in Section 2.1 applied to the action of the VSUSY generators on the spinorial fields given in Table \ref{tablespinning} lead to the conclusion that this is not possible in this case.


\subsection{Another $Y=0$ multiplet}

Inspired by Refs. \cite{Kato:2005gb} and \cite{Kato:2008dw}, we construct a multiplet with the same field content as in the previous section but with a nontrivial action of the central charge operators on the fields.\\
We are interested in multiplets satisfying (\ref{pzz}).
For simplicity, we work in the case $Z=-\tilde{Z}$\footnote{This is the same kind of restriction as choosing to have a real central charge instead of a complex one in $\mathcal{N}=2$ standard SUSY. This choice is made for instance in the $\mathcal{N}=2$ SUSY and related twisted SUSY papers we are inspired by in this paper, like \cite{Kato:2005gb} \cite{Kato:2008dw} and \cite{Sohnius:1978fw}.}.
Since the central charges are real, the case $Z= \tilde{Z}$ corresponds to fields with an imaginary mass (tachyons) and for this reason we do not consider it.\\
As in the multiplet presented in the previous section, the field content is four Dirac spinors, $\psi$, $\tilde\chi$, $\lambda$, $\tilde\xi$, two even and two odd. They transform as in Table \ref{y0}.
\begin{table}[h]
\begin{center}
 \begin{tabular}{|c|c|c|c|}
    \hline
      & $Q_5$ & $Q_{\mu}$ & $Z$\\
     \hline
      & & &\\
    $\psi$ & $-\tilde{\chi}$ & $-\gamma_{\mu}\tilde{\xi}$ & $i\lambda$\\
    $\tilde{\chi}$ & $-\frac{i}{2}\lambda$ & $-\frac{i}{2}\dslash \gamma_{\mu} \psi$ & $  -i\dslash\tilde{\xi}$\\
    $\lambda$ & $\dslash\tilde{\xi}$ & $-\gamma_{\mu}\dslash \tilde{\chi }$ & $  -i\Box\psi$\\
    $\tilde{\xi}$ & $-\frac{i}{2} \dslash\psi$ & $\frac{i}{2}\gamma_{\mu}\lambda$ & $i\dslash\tilde{\chi}$ \\
     & & &\\
    \hline
    \end{tabular}
    \end{center}
\caption{Action of the VSUSY charges on the fields (four-spinor multiplet with nontrivial action of the central charge). \label{y0}}
\end{table}

\noindent One can check that this representation indeed satisfies the constraint $P^{2}=-Z^{2}$. Since this constraint allows for the presence of the odd Casimir but there is no natural odd constant, the odd Casimir vanishes on all fields.
The transformations of the barred field are given in Table \ref{tablespinorsbar2}.\\
\begin{table}[h!]
 \begin{center}
\begin{tabular}{|c|c|c|c|}
    \hline
      & $\delta_{Q_5}$ & $\delta_{Q_{\mu}}$ & $\delta_{Z}$\\
     \hline
      & & &\\
    $\bar\psi$ & $i\bar{\tilde\chi}\e_5$ & $-i\bar{\tilde{\xi}}\gamma_\m \e^\m$ & $-\bar\lambda\e_Z$\\
    $\bar{\tilde\chi}$ & $\frac{1}{2}\bar\lambda \e_5$ & $\frac{1}{2}\bar\psi \gamma_\m \overleftarrow{\dslash}\e^\m$ & $-\bar{\tilde{\xi}}\overleftarrow{\dslash}\e_Z  $\\
    $\bar\lambda$ & $i\bar{\tilde\xi}\overleftarrow{\dslash}\e_5$ & $i\bar{\tilde\chi}\overleftarrow{\dslash}\gamma_\m \e^\m$ & $\bar\psi\overleftarrow{\Box} \e_Z$\\
    $\bar{\tilde{\xi}}$ & $-\frac{1}{2}\bar\psi \overleftarrow{\dslash}\e_5$ & $\frac{1}{2}\bar\lambda\gamma_\m \e^\m$ &  $\bar{\tilde\chi}\overleftarrow{\dslash}\e_Z$ \\
     & & &\\
    \hline
    \end{tabular}
\end{center}
\caption{Transformations of the Dirac-conjugate spinor fields (four-spinor multiplet with nontrivial action of the central charge) \label{tablespinorsbar2}}
\end{table}
\vskip 12pt
\noindent One can check that the following Klein-Gordon type free action is invariant under VSUSY:
\begin{equation}
 S=\int d^4 x~ \left(\bar{\tilde{\xi}}\Box \tilde{\xi} + \bar{\tilde\chi}\Box \tilde{\chi} +\frac{i}{2}\bar\psi \Box \lambda - \frac{i}{2}\bar \lambda \Box \psi+ m^2( \bar{\tilde{\xi}} \tilde{\xi} + \bar{\tilde\chi} \tilde{\chi} +\frac{i}{2}\bar\psi \lambda - \frac{i}{2}\bar \lambda \psi)\right).
\label{action4}
\end{equation}
Following the discussion at the end of Section 2.1, we see that for this multiplet it is possible to impose the Majorana condition (\ref{majorana}) on the spinors to reduce the number of degrees of freedom of the representation by half, by requiring all of them to be `real' in the sense stated there.\\
A comment is due on the unexpected form of the action. It is of course unnatural to have an action for spinor fields with Klein-Gordon kinetic terms.
A first check is that a set of Dirac-like equations for the spinor fields indeed cannot be invariant under the VSUSY transformations given in Table \ref{y0}. The puzzling issue is however that, by inspection of action (\ref{action4}) only, one cannot see that the Lorentz group under which the spinors transform is the same as the spacetime Lorentz group, so apparently the symmetry group of the action is larger than the VSUSY algebra (\ref{VSUSYalgebra}). This feature is related to the fact that no spinor generators are present in the VSUSY algebra, so spinors are introduced by hand in the representations and do not naturally arise as in the case of ordinary SUSY as superpartners of scalar or vectorial fields. However, the VSUSY transformations under which the action is invariant do show that the two Lorentz groups have to be identified. This puzzle is technically due to the absence of a coupling between $\gamma_\m$ and $\partial_\m$ in the kinetic terms for the spinors and could be solved by possible VSUSY-invariant interaction terms where this coupling would be present. This issue deserves further investigation. The search for a VSUSY invariant interacting field theory involving this multiplet is left for future work.

\subsubsection{On-shell $Y=0$ multiplet}
It is also possible to construct a multiplet with only two spinors, one even and one odd, being a realization of the irreducible doublet with $Y=0$ discussed in Section 2.2.\\ We work again in the simpler case $Z=-\tilde{Z}$.\\
Consider the multiplet given in Table \ref{tableonshell}, where $\hat z$ is the value of the central charge.
\begin{table}[h]
\begin{center}
\begin{tabular}{|c|c|c|}
    \hline
      & $Q_5$ & $Q_{\mu}$\\
     \hline
     & &\\
    $\psi$ & $-\tilde{\chi}$ & $-\frac{i}{\hat z}\gamma_{\mu}\dslash\tilde{\chi}$\\
    $\tilde{\chi}$ & $-\frac{\hat z}{2}\psi$ & $-\frac{i}{2}\dslash\gamma_{\mu}\psi$\\
     & &\\
    \hline
    \end{tabular}
\end{center}
\caption{Action of the VSUSY charges on the fields (on-shell two-spinor multiplet).\label{tableonshell}}
\end{table}
The transformations of the barred fields are given in Table \ref{tableonshell2}.
\begin{table}[h]
\begin{center}
\begin{tabular}{|c|c|c|}
    \hline
      & $\delta_{Q_5}$ & $\delta_{Q_{\mu}}$\\
     \hline
     & &\\
    $\bar\psi$ & $i\bar{\tilde\chi}\e$ & $\frac{1}{\hat z}\bar{\tilde\chi}\overleftarrow{\dslash}\gamma_{\mu}\e^{\mu}$\\
    $\bar{\tilde{\chi}}$ & $\frac{i\hat z}{2}\bar\psi\e$ & $\frac{1}{2}\bar\psi\gamma_{\mu}\overleftarrow{\dslash}\e^{\mu}$\\
     & &\\
    \hline
    \end{tabular}
\end{center}
\caption{Transformations of the Dirac-conjugate spinor fields (on-shell two-spinor multiplet).\label{tableonshell2}}
\end{table}
One can easily see that this table is a component realization of the VSUSY algebra (\ref{VSUSYalgebra}) in the special case $Z=-\tilde{Z}$ when one imposes the equations of motion $\Box \psi=\hat z^2 \psi$ and $\Box \tilde{\chi}=\hat z^2 \tilde{\chi}$, which are a realization of the constraint $P^2=-Z^2$ on the fields.
Note also that the odd Casimir is realized on this multiplet with the value zero.

\noindent Actually, this short multiplet can be derived from the one presented in Table \ref{y0} by requiring that the $Z$ operator acts as a number $\hat z$ on all fields.
The resulting four constraint equations amount to 
\begin{equation}
\lambda=-i\hat z\psi\,, \quad\quad\quad\quad \tilde\xi=\frac{i}{\hat z} \dslash \tilde\chi\,,
\label{reduction}
\end{equation}
reducing the number of fields from four to two, together with the dynamical equations
\begin{equation}
 \Box \psi= \hat z^2 \psi\,,  \quad\quad\quad\quad \Box \tilde \xi= \hat z^2 \tilde \xi\,.
\end{equation}
Therefore the truncation procedure puts the remaining fields on-shell and so it is not correct to derive an action for the new multiplet by directly applying this truncation to action (\ref{action4}).
However, there exists a VSUSY invariant action for this shorter multiplet:
\begin{equation}
 S=\int d^4 x~ \Big(2 \bar{\tilde\chi}\Box \tilde{\chi} + \hat z\bar \psi \Box \psi+ m^2(2 \bar{\tilde\chi} \tilde{\chi} +\hat z\bar \psi \psi)\Big).
\end{equation}
Concerning the possibility of reducing the number of degrees of freedom of this multiplet by imposing a Majorana condition on the spinors, inspection of constraints (\ref{reduction}) shows that these are not compatible with such condition and therefore the number of degrees of freedom cannot be reduced further.
This can be also seen directly at the level of the action since, for Majorana spinors, the terms involving the field $\psi$ in the action would vanish.

\subsection{$Y=0$ multiplets from superspace a la Sohnius}
In this section we would like to derive the spinor multiplets presented in Tables \ref{tablespinning}, \ref{y0} and \ref{tableonshell} via a superspace approach.\\
Since the multiplet in Table \ref{y0} is characterized by a nontrivial action of the central charge on the component fields, it will not be possible to derive it by using a superspace of the standard type, spanned only by the bosonic coordinates $x^\m$ associated to the generator $P^\m$ and the fermionic coordinates $\theta^A$ associated to the fermionic generators $Q^A$. Two extra bosonic coordinates $z$ and $\tilde z$ associated to the central charge operators $Z$ and $\tilde Z$ must be present as well. For simplicity, we consider the case where $\tilde Z=-Z$ so that we have to add only one extra bosonic coordinate $z$. In contrast to the case of extra fermionic coordinates, automatically leading to a superfield expansion with a finite number of components, the presence of an extra bosonic coordinate leads to an infinite number of component fields. To obtain a multiplet with a finite number of fields, one must necessarily impose a set of covariant constraints in superspace forcing the coefficients of the $z$ expansion for higher powers of $z$ to be functions of a finite number of lower coefficients.\\
This procedure has been discussed first by Sohnius in \cite{Sohnius:1978fw}, in the context of $\mathcal{N}=2$ supersymmetric theories with one real central charge. There he introduced a superspace with supercoordinates  $(x^{\mu},\theta^{\alpha}_{i}, \overline{\theta}^{\dot\alpha i}, z)$, where $\alpha$ and $\dot\alpha$ are spinor indices and $i$ is the internal $SU(2)$ R-symmetry index. Starting from the most general expansion of a superfield $\Phi_{i}(x^{\mu},\theta^{\alpha}_{i}, \overline{\theta}^{\dot\alpha i}, z)$ with an extra $SU(2)$ index $i$, he imposed two covariant constraints on it in order to reduce the number of degrees of freedom. In practice, he imposed that the covariant spinor derivatives $D_{\alpha}^{j}$ and $\overline{D}_{\dot\alpha j}$, when acting on $\Phi_{i}$, produce something proportional to the only two structures with two indices available in the $SU(2)$ space, the Kronecker delta $\delta^{\,j}_{i}$ and the antisymmetric symbol $\e_{ij}$:
\begin{equation}
D_{\alpha}^{j}\Phi_{i} = \delta^{\,j}_{i} \Psi_{\alpha}\,,\quad \quad \quad \quad \overline{D}_{\dot\alpha j}\Phi_{i}=\e_{ij}\overline{\Psi}_{\dot\alpha}\,.
\label{sohnius}
\end{equation}
In order for (\ref{sohnius}) to be covariant constraints, $\Psi_{\alpha}$ and $\overline{\Psi}_{\dot\alpha}$ must be full superfields with a spinor index. Moreover, these two constraints automatically imply that $P^2~\sim~Z^2$ on $\Phi_{i}$, which means that the superfield satisfies $\Box \Phi_{i}=\frac{\partial^{2}}{\partial z^{2}}\Phi_{i}$. As a result, the range in spin covered by the supermultiplet is only $\Delta s=\frac{1}{2}$. This ensures that the supermultiplet contains only four independent component fields ($A$, $\psi$, $\overline{\varphi}$, $F$) and that the higher order terms in the series expansion in $z$ are simply higher derivatives of those fields.

We would like to follow this approach to construct our VSUSY spinor multiplets. However, as we will discuss in detail in the next section, in our case we have an odd vectorial superspace covariant derivative, $D_{\mu}$, and an odd scalar one, $D_5$, both carrying no R-symmetry indices. Since we want to derive multiplets containing only spinor components, but no spinor indices appear in the VSUSY algebra (\ref{VSUSYalgebra}), we must necessarily start from a superfield carrying a spinor index $\Phi^{\alpha}$.
Moreover, since the only structure containing both vector and spinor indices at our disposal is the Dirac $(\gamma_{\mu})^{\alpha}_{\,\,\beta}$, a natural covariant constraint to be imposed on $\Phi^{\alpha}$ is
\begin{equation}
D_{\mu}\Phi^{\alpha} = (\gamma_{\mu})^{\alpha}_{\,\,\beta}\tilde\Lambda^{\beta}\,,
\label{first}
\end{equation}
where $D_{\mu}$ is the odd vectorial superspace covariant derivative, to be explicitly given in the next section, and $\tilde\Lambda^{\alpha}$ is a generic spinor superfield.\\
As we will show in the following, this constraint indeed does the job of eliminating all higher spin components in the superfield, leaving only spinor components in the superfield expansion.
However, this constraint by itself is not enough to ``terminate'' the expansion in $z$. As much as Sohnius needed two contraints, one involving $D_{\alpha}^{j}$ and the other $\overline{D}_{\dot\alpha j}$, to obtain a finite number of components fields, we will also need to impose a second constraint on $\Phi^{\alpha}$ containing the covariant derivative $D_{5}$ in order to obtain a finite number of component fields. This second constraint will have the following property: combined with (\ref{first}) it will imply that $P^2=-Z^2$ on $\Phi^{\alpha}$, ensuring that the infinite series expansion in $z$ actually depends only on a finite number of lower components.
We will see that the right choice for a second constraint is of the form
\begin{equation}
\Dslash D_{5}\Phi^{\alpha} \sim \dslash\Phi^{\alpha}\,.
\label{second}
\end{equation}
In principle, one could also think of a constraint of the form $D_5 \Phi \sim \tilde \Lambda$. In fact, one can check that this also implies that $P^2=-Z^2$ on $\Phi^\a$. However, inspection of the component constraint equations one gets from it shows that it is stronger than (\ref{second}), in fact too strong, since it directly implies equations of motions for the component fields and not constraints reducing the number of components.

\subsubsection{Superspace}
We consider a superspace with supercoordinates $x^A\equiv(x^{\mu},\theta^{\mu}, \theta_{5}, z, \tilde z)$.\\
The VSUSY charges have the following form:
\begin{equation}
Q_{\mu}= \frac{\partial}{\partial \theta^{\mu}}-\frac{i}{2}\theta_{5}\frac{\partial}{\partial x^{\mu}}+\frac{i}{2}\theta_{\mu}\frac{\partial}{\partial z}\,, \quad \quad \quad Q_{5}= \frac{\partial}{\partial \theta_{5}}-\frac{i}{2}\theta_{\mu}\frac{\partial}{\partial x^{\mu}}+\frac{i}{2}\theta_{5}\frac{\partial}{\partial \tilde z}\,.
\label{chargessuper}
\end{equation}
They satisfy the VSUSY algebra:
\begin{equation}
\{Q_{\mu},Q_{\nu}\}=\eta_{\mu\nu}Z\,, \quad\quad\quad \{Q_{\mu},Q_{5}\}=-P_{\mu}\,,\quad\quad\quad \{Q_{5},Q_{5}\}=\tilde Z\,.
\label{algebra}
\end{equation}
where
\begin{equation}
Z= i\frac{\partial}{\partial z}\,, \quad\quad \tilde Z = i \frac{\partial}{\partial \tilde z}\,,\quad \quad P_{\mu}=i\frac{\partial}{\partial x^{\mu}}\,.
\end{equation}
The superspace covariant derivatives are\footnote{Due to the fact that in VSUSY superspace both the even coordinate $x^\m$ and the odd coordinate $\theta^\m$ carry a vectorial index, the notation for the covariant derivatives could lead to some misunderstanding. We choose to denote with $D_\m$ the odd vectorial covariant derivative and simply with $\frac{\partial}{\partial x^\m}\equiv\partial_\m$ the even one.}
\begin{equation}
D_A=\left(\frac{\partial}{\partial x^\m}, D_\n, D_5, \frac{\partial}{\partial z}, \frac{\partial}{\partial \tilde z}\right),
\label{covdevI}
\end{equation}
where $D_\m$ and $D_5$ are given by
\begin{equation}
D_{\mu}= \frac{\partial}{\partial \theta^{\mu}}+\frac{i}{2}\theta_{5}\frac{\partial}{\partial x^{\mu}}-\frac{i}{2}\theta_{\mu}\frac{\partial}{\partial z}\,, \quad \quad \quad D_{5}= \frac{\partial}{\partial \theta_{5}}+\frac{i}{2}\theta_{\mu}\frac{\partial}{\partial x^{\mu}}-\frac{i}{2}\theta_{5}\frac{\partial}{\partial \tilde z}\,.
\label{covdevII}
\end{equation}
It is easy to see that these anticommute with the VSUSY charges and satisfy:
\begin{equation}
\{D_{\mu},D_{\nu}\}=-\eta_{\mu\nu}Z\,, \quad\quad\quad \{D_{\mu},D_{5}\}=P_{\mu}\,,\quad\quad\quad \{D_{5},D_{5}\}=-\tilde Z\,.
\end{equation}
Note that, similarly as in ordinary superspace, one can perform the following change of variables
\begin{equation}
x^{\m(\pm)}=x^\m\pm \frac{i}{2} \theta^\m \theta_5\,,
\label{covdertrans}
\end{equation}
where the other supercoordinates remain unchanged,
so that the odd covariant derivatives take the asymmetric form
\begin{equation}
D^{(+)}_\m= \frac{\partial}{\partial \theta^{\mu}}+i\theta_{5}\frac{\partial}{\partial x^{\mu}}-\frac{i}{2}\theta_{\mu}\frac{\partial}{\partial z}\,,\quad\quad
D^{(+)}_5=\frac{\partial}{\partial \theta_{5}}-\frac{i}{2}\theta_{5}\frac{\partial}{\partial \tilde z}\,.
\label{newcovderivI}
\end{equation}
or
\begin{equation}
D^{(-)}_\m= \frac{\partial}{\partial \theta^{\mu}}-\frac{i}{2}\theta_{\mu}\frac{\partial}{\partial z}\,,\quad\quad D^{(-)}_5=\frac{\partial}{\partial \theta_{5}}+i\theta_{\mu}\frac{\partial}{\partial x^{\mu}}-\frac{i}{2}\theta_{5}\frac{\partial}{\partial \tilde z}\,.
\label{newcovderivII}
\end{equation}
This is the VSUSY analogue of the chiral and antichiral superspace representations vs. the vector representation of ordinary supersymmetry.\\
In the following we will write constraints in superspace to reduce the number of degrees of freedom of a general superfield. We remind the reader that all constraints written in terms of the superspace covariant derivatives (\ref{covdevI}), (\ref{covdevII}) and in terms of superfields are automatically VSUSY invariant.
\subsubsection{Spinor Superfield}
For simplicity, let us first restrict ourselves to the case $\tilde Z=-Z$.\\
As said before, we expect to be able to obtain a supermultiplet containing only spinor components, so we choose to consider a superfield with an extra spinor index $\Phi^{\alpha}$ expanded in $\theta^{\mu}$, $\theta_{5}$ and $z$. We also choose $\Phi^{\alpha}$ to be bosonic (as usual, we put a tilde on its fermionic components):
\begin{eqnarray}
\Phi^{\alpha}(x^{\mu},\theta^{\mu},\theta_{5},z)&=&\phi^\a(x,z)+
\tilde{\phi}^\a(x,z)\theta_{5}
+\tilde\psi_{\mu}^\a(x,z)\theta^{\mu}+
\psi_{\mu}^\a(x,z)\theta^{\mu}\theta_{5}\cr
&&+\xi_{\mu\nu}^\a(x,z)\theta^{\mu}\theta^{\nu}+
\tilde{\xi}_{\mu\nu}^\a(x,z)\theta^{\mu}\theta^{\nu}\theta_{5}
+\tilde\chi_{\mu\nu\rho}^\a(x,z)\theta^{\mu}\theta^{\nu}
\theta^{\rho}\cr
&&+\chi_{\mu\nu\rho}^\a(x,z)
\theta^{\mu}\theta^{\nu}\theta^{\rho}\theta_{5}
+\zeta_{\mu\nu\rho\sigma}^\a(x,z)
\theta^{\mu}\theta^{\nu}\theta^{\rho}\theta^{\sigma}+
\tilde{\zeta}_{\mu\nu\rho\sigma}^\a(x,z)
\theta^{\mu}\theta^{\nu}\theta^{\rho}\theta^{\sigma}\theta_{5}.\cr
&&
\label{spinorsuperfield}
\end{eqnarray}
where we assume that every component field admits a Taylor expansion in $z$, for instance
\begin{equation}
\phi^\a(x,z)=\sum_{n=0}^{\infty}\frac{1}{n!}z^{n}\,\phi^{\a(n)}(x)
\label{zetaexp}
\end{equation}
and similar for the other components.\\
Since the $\theta$'s anticommute, it is clear that $\xi_{\mu\nu}$, $\tilde{\xi}_{\mu\nu}$, $\chi_{\mu\nu\rho}$, $\tilde{\chi}_{\mu\nu\rho}$, $\zeta_{\mu\nu\rho\sigma}$ and $\tilde{\zeta}_{\mu\nu\rho\sigma}$ must be totally antisymmetric in their vectorial indices.\\
In order to avoid a cumbersome notation, we will often omit the spinor indices on the components.
\subsubsection{Supercovariant constraints}
 Saying that the constraint (\ref{first}) does the job of removing all higher-spin Rarita-Schwinger-like fields and leaving only spinors components in the expansion (\ref{spinorsuperfield}), can technically be re-expressed by saying that the components ($\psi_\m$, $\tilde\psi_\m$, $\xi_{\mu\nu}$, $\tilde{\xi}_{\mu\nu}$, $\chi_{\mu\nu\rho}$, $\tilde{\chi}_{\mu\nu\rho}$, $\zeta_{\mu\nu\rho\sigma}$, $\tilde{\zeta}_{\mu\nu\rho\sigma}$) solving the constraint only contain the following two structures
\begin{equation}
A^{\alpha}_{\mu_{1}\mu_{2}...\mu_{n}}(x)=\gamma_{\mu_{1}\mu_{2}...\mu_{n}}B^{\alpha}(x)+
\gamma_{[\mu_{1}\mu_{2}...\mu_{n-1}}\partial_{\mu_{n}]}C^{\alpha}(x)\,,
\label{forms}
\end{equation}
where
\begin{equation}
\gamma_{\mu_{1}\mu_{2}...\mu_{n}}=\frac{1}{n!}\sum_{perm} (-)^{\sigma(\mu_{1}...\mu_{n})}\, \gamma_{\mu_{1}}\gamma_{\mu_{2}}...\gamma_{\mu_{n}}
\end{equation}
and with $\sigma(\mu_{1}...\mu_{n})$ we denote a permutation of the indices $\mu_{1}...\mu_{n}$.
As shown in Appendix B, the general solution of constraint (\ref{first}) is given by the following relations for $n=0,1,2,...$:
\begin{eqnarray}
&\tilde\psi_{\mu}^{(n)}=\gamma_{\mu} \tilde\lambda^{(n)}\quad\quad\quad &\psi_{\mu}^{(n)}=\gamma_{\mu} \lambda^{(n)}-\frac{i}{2}\partial_{\mu}\phi^{(n)}\nonumber\\
&\xi_{\mu\nu}^{(n)}=-\frac{i}{4}\gamma_{\mu\nu} \phi^{(n+1)}\quad\quad\quad &\tilde\xi_{\mu\nu}^{(n)}=-\frac{i}{4}\gamma_{\mu\nu} \tilde\phi^{(n+1)}-\frac{i}{2}\gamma_{[\mu}\partial_{\nu]}\tilde\lambda^{(n)}\nonumber\\
&\tilde\chi_{\mu\nu\rho}^{(n)}=-\frac{i}{12}\gamma_{\mu\nu\rho} \tilde\lambda^{(n+1)}\quad\quad\quad &\chi_{\mu\nu\rho}^{(n)}=-\frac{i}{12}\gamma_{\mu\nu\rho} \lambda^{(n+1)}-\frac{1}{8}\gamma_{[\mu\nu}\partial_{\rho]}\phi^{(n+1)}\nonumber\\
&\zeta_{\mu\nu\rho\sigma}^{(n)}=-\frac{1}{96}\gamma_{\mu\nu\rho\sigma} \phi^{(n+2)}\quad\quad\quad &\tilde\zeta_{\mu\nu\rho\sigma}^{(n)}=-\frac{1}{96}\gamma_{\mu\nu\rho\sigma} \tilde\phi^{(n+2)}-\frac{1}{24}\gamma_{[\mu\nu\rho}\partial_{\sigma]}\tilde\lambda^{(n+1)}.
\label{exp}
\end{eqnarray}
This is indeed of the expected form (\ref{forms}).\\
We are left with four infinite sets of independent field components,
\begin{equation}
\phi^{(n)}, \quad\quad\tilde\phi^{(n)},\quad\quad\lambda^{(n)},\quad\quad\tilde\lambda^{(n)},
\label{infty}
\end{equation}
whose VSUSY transformations are given in Table \ref{table}.\\
\begin{table}[h]
\begin{center}
\begin{tabular}{|c|c|c|c|}
    \hline
      & $Q_{5}$ & $Q_{\mu}$ & $Z$\\
     \hline
      & & &\\
    $\phi^{(n)}$ & $-\tilde\phi^{(n)}$ & $-\gamma_{\mu}\tilde\lambda^{(n)}$ & $i\,\phi^{(n+1)}$\\
    $\tilde\phi^{(n)}$ & $-\frac{i}{2}\phi^{(n+1)}$ & $\gamma_{\mu}\lambda^{(n)}-i\partial_{\mu}\phi^{(n)}$ & $i\, \tilde\phi^{(n+1)}$ \\
    $\tilde\lambda^{(n)}$ & $-\lambda^{(n)}$ & $\frac{i}{2}\gamma_{\mu}\phi^{(n+1)}$ & $i\,\tilde\lambda^{(n+1)}$\\
    $\lambda^{(n)}$ & $-\frac{i}{2}\tilde\lambda^{(n+1)}$ & $\,\,\,\,\,\,-\frac{i}{2}\gamma_{\mu}\tilde\phi^{(n+1)}-i\partial_{\mu}\tilde\lambda^{(n)}$ & $i\, \lambda^{(n+1)}$ \\
    & & &\\
    \hline
    \end{tabular}
\end{center}
\caption{Action of the VSUSY charges on the four infinite sets of spinor components (\ref{infty}) solving constraint (\ref{first}). \label{table}}
\end{table}

If we want to recover the case with $Z$ acting like a number we need to further impose the constraint
\begin{equation}
Z\Phi=\hat{z}\,\Phi\,,
\label{zeta}
\end{equation}
with $\hat z$ number. This constraint implies:
\begin{equation}
\phi^{(n+1)}=-i\hat z\phi^{(n)}\,,\quad \quad \tilde\phi^{(n+1)}=-i\hat z\tilde\phi^{(n)}\,,\quad \quad \lambda^{(n+1)}=-i\hat z\lambda^{(n)}\,,\quad \quad \tilde \lambda^{(n+1)}=-i\hat z\tilde\lambda^{(n)}\,.
\label{sup}
\end{equation}
We are then left with only four independent field components, $\phi^{(0)}$, $\tilde\phi^{(0)}$, $\lambda^{(0)}$ and $\tilde\lambda^{(0)}$, whose VSUSY transformations are given in Table \ref{tablenine}.
\begin{table}[h]
\begin{center}
\begin{tabular}{|c|c|c|}
    \hline
      & $Q_{5}$ & $Q_{\mu}$\\
     \hline
      & & \\
    $\phi^{(0)}$ & $-\tilde\phi^{(0)}$ & $-\gamma_{\mu}\tilde\lambda^{(0)}$ \\
    $\tilde\phi^{(0)}$ & $-\frac{\hat z}{2}\phi^{(0)}$ & $\gamma_{\mu}\lambda^{(0)}-i\partial_{\mu}\phi^{(0)}$ \\
    $\tilde\lambda^{(0)}$ & $-\lambda^{(0)}$ & $\frac{\hat z}{2}\gamma_{\mu}\phi^{(0)}$ \\
    $\lambda^{(0)}$ & $-\frac{\hat z}{2}\tilde\lambda^{(0)}$ & $\,\,\,\,\,\,-\frac{\hat z}{2}\gamma_{\mu}\tilde\phi^{(0)}-i\partial_{\mu}\tilde\lambda^{(0)}$\\
    & & \\
    \hline
    \end{tabular}
\end{center}
\caption{Action of the VSUSY charges on the four spinor multiplet obtained from the components given in Table \ref{table} by further diagonalizing the central charge. \label{tablenine}}
\end{table}

\noindent One can check that these transformations are not compatible with the Dirac equation.\\
However, it is easy to show that the following linear combinations:
\begin{eqnarray}
&\psi_{1}\equiv-\frac{i}{2}(\phi^{(0)}+\frac{2i}{\hat z}\lambda^{(0)})\,; \quad \quad \psi_{2}\equiv\frac{1}{2}\gamma_5(\phi^{(0)}-\frac{2i}{\hat z}\lambda^{(0)})\,;\nonumber\\
&\tilde \psi_{1}\equiv\frac{1}{\sqrt{2\hat z}}\gamma_5(\tilde\phi^{(0)}-i\tilde\lambda^{(0)})\,;\quad\quad\tilde \psi_{2}\equiv-\frac{i}{\sqrt{2\hat z}}(\tilde\phi^{(0)}+i\tilde\lambda^{(0)})\,,
\label{lincomb}
\end{eqnarray}
exactly reproduce the multiplet previously given in Table \ref{tablespinning} with the following identifications:
\begin{equation}
\hat z=m=a ~~~~~~(Z=-\tilde Z)\,.
\end{equation}
Now we go back to the supermultiplet (\ref{infty}) where the operator $Z$ is still not diagonal and we want to
find a second constraint that, together with the first given in (\ref{first}), automatically implies the multiplet shortening conditions $P^2=-Z^2$.
As already mentioned in the introduction to this section, a natural choice is:
\begin{equation}
\dslash\Phi=-\frac{i}{2}D_{5}\Dslash\,\Phi\,,
\end{equation}
or, equivalently,
\begin{equation}
\dslash \,\Phi=-2i\,D_{5}\tilde\Lambda\,,
\end{equation}
where $\tilde\Lambda$ is the spinor superfield appearing in (\ref{first}). \\
After some algebraic manipulations one can prove that (\ref{second}), combined with (\ref{first}), implies that $P^{2}=-Z^{2}$, that is to say:
\begin{equation}
\Box\,\Phi^{\alpha}=-\frac{\partial^{2}}{\partial z^{2}}\Phi^{\alpha}\, .
\label{box}
\end{equation}
The constraint (\ref{box}) reduces the number of independent components down to eight:
\begin{equation}
\phi^{(0)}, \quad \phi^{(1)},\quad\tilde\phi^{(0)},\quad\tilde\phi^{(1)},\quad\lambda^{(0)}, \quad \lambda^{(1)},\quad\tilde\lambda^{(0)},\quad\tilde\lambda^{(1)}.
\end{equation}
Higher order components (for $n=1,2...$) are related to the previous ones via:
\begin{eqnarray}
&\phi^{(2n)}=(-\Box)^{n}\phi^{(0)}\quad\quad&\tilde\phi^{(2n)}=(-\Box)^{n}\tilde\phi^{(0)}\nonumber\\
&\lambda^{(2n)}=(-\Box)^{n}\lambda^{(0)}\quad\quad&\tilde\lambda^{(2n)}=(-\Box)^{n}\tilde\lambda^{(0)}\nonumber\\
&\phi^{(2n+1)}=(-\Box)^{n}\phi^{(1)}\quad\quad&\tilde\phi^{(2n+1)}=(-\Box)^{n}\tilde\phi^{(1)}\nonumber\\
&\lambda^{(2n+1)}=(-\Box)^{n}\lambda^{(1)}\quad\quad&\tilde\lambda^{(2n+1)}=(-\Box)^{n}\tilde\lambda^{(1)}.
\label{higher}
\end{eqnarray}
Now, if we implement (\ref{second}), which is stronger than (\ref{box}), we obtain further constraints on the component fields:
\begin{equation}
\lambda^{(0)}=\frac{i}{2}\dslash \phi^{(0)}\,, \quad \quad \lambda^{(1)}=\frac{i}{2}\dslash \phi^{(1)}\,, \quad \quad \tilde\lambda^{(1)}=\,\dslash\tilde\phi^{(0)}\,,\quad \quad \tilde \phi^{(1)}=-\dslash\tilde\lambda^{(0)}.
\end{equation}
So, by imposing both (\ref{first}) and (\ref{second}), one reduces the number of independent components down to four:
\begin{equation}
\phi^{(0)}, \quad \quad \phi^{(1)},\quad\quad\tilde\phi^{(0)},\quad\quad \tilde\lambda^{(0)}.
\label{comp}
\end{equation}
The superfield expansion in $\theta^{\mu}$, $\theta_{5}$ and $z$ reads:
\begin{eqnarray}
\Phi^{\alpha}=&\phi^{(0)} + \tilde{\phi}^{(0)}\theta_{5}+\gamma_{\mu}\tilde\lambda^{(0)}\theta^{\mu}+z\,\phi^{(1)}- z\,\dslash\tilde\lambda^{(0)}\theta_{5}+z\,\gamma_{\mu}\dslash\tilde\phi^{(0)}\theta^{\mu}
+\frac{i}{2}\gamma_{\mu}\dslash\phi^{(0)}\theta^{\mu}\theta_{5}+\nonumber\\
&- \frac{i}{2}\partial_{\mu}\phi^{(0)}\theta^{\mu}\theta_{5} - \frac{i}{4} \gamma_{\mu\nu}
\phi^{(1)}\theta^{\mu}\theta^{\nu} -\frac{1}{2}z^{2}\,\Box\,\phi^{(0)}+{\rm higher\,\, order\,\, terms},
\end{eqnarray}
where the higher order terms are at least trilinear in $\theta^{\mu}$, $\theta_{5}$ and $z$ and contain only the fields appearing in (\ref{comp}).\\
The VSUSY transformations for this supermultiplet are the ones already given in Table \ref{y0}, once one makes the following identification:
\begin{equation}
\phi^{(0)}\equiv\psi\,, \quad \quad \tilde{\phi}^{(0)}\equiv\tilde\chi\,, \quad \quad \phi^{(1)}\equiv\lambda\,,
\quad \quad \tilde{\lambda}^{(0)}\equiv\xi\,.
\end{equation}
Now, if we diagonalize $Z$ as we did before, we get the following relations:
\begin{equation}
\phi^{(1)}=-i\hat z \phi^{(0)} \quad\quad\quad\quad \tilde\lambda^{(0)}=\frac{i}{\hat z} \dslash\tilde\phi^{(0)}.
\end{equation}
So we are left with the multiplet of Table \ref{tabletwofields}, containing just two fields, which closes by imposing the constraint $P^2=-Z^2$ by hand.
\begin{table}[h]
\begin{center}
\begin{tabular}{|c|c|c|}
    \hline
      & $Q_{5}$ & $Q_{\mu}$\\
     \hline
      & & \\
    $\phi^{(0)}$ & $-\tilde\phi^{(0)}$ & $-\frac{i}{\hat z} \gamma_{\mu}\dslash\tilde\phi^{(0)}$ \\
    $\tilde\phi^{(0)}$ & $-\frac{\hat z}{2}\phi^{(0)}$ & $-\frac{i}{2}\dslash\gamma_{\mu}\phi^{(0)}$ \\
    & & \\
    \hline
    \end{tabular}
\end{center}
\caption{Action of the VSUSY charges on the two spinor fields obtained from the Table \ref{tablenine} by further imposing the constraint (\ref{second}) \label{tabletwofields}}
\end{table}

\noindent One can check that the previous table reduces to Table \ref{tableonshell} when the following identification is implemented:
\begin{equation}
\tilde\phi^{(0)}= \tilde\chi\,, \quad\quad\quad\quad \phi^{(0)}=\psi\,.
\end{equation}

\section{$Y=1/2$ multiplet and twisted topological models}
In \cite{Kato:2005gb} and \cite{Kato:2008dw}, an ${\mathcal N}=2$ twisted superspace formalism with a central charge in
four-dimensional Euclidean space is constructed, by introducing a Dirac-K\"ahler twist.
This twist leads to the following supertranslation algebra
\begin{eqnarray}
&&\{Q_{\mu},Q_{\nu}\}=Z \delta_{\mu\nu}\,,\quad \quad \{Q_{5},Q_{5}\}=Z\,,\quad  \{Q_{\mu},Q_{5}\}=-P_{\mu},\nonumber\\
&&\{Q^+_{\m\n},Q_5\}=0\,,\quad
\{Q^+_{\m\n},Q_\rho\}=\delta^+_{\m\n,\rho\s}P^\s\,,\quad
\{Q^+_{\m\n},Q^+_{\rho\s}\}=\delta^+_{\m\n,\rho\s}Z\,,
 \label{TSUSY}
\end{eqnarray}
 which is clearly an extension of the supertranslation sector of the VSUSY algebra (\ref{VSUSYalgebra}) in the case where the two VSUSY central charges are
 identified. \
In (\ref{TSUSY}) $Q^{+}_{\m\n}$ denotes an antiselfdual two-form in Euclidean space and $\delta^+_{\m\n,\rho\s}=\delta_{\m\rho}\delta_{\n\s} -\delta_{\n\rho}\delta_{\m\s}-\e_{\m\n\rho\s}$ is, up to a constant, a projector on the selfdual space.
We will call this algebra tensor supersymmetry algebra (TSUSY), due to the fact that the extra odd generator present there with respect to VSUSY is an antiselfdual two-form $Q^{+}_{\m\n}$.\\
In \cite{Kato:2005gb} and \cite{Kato:2008dw}, multiplets which are a representation of TSUSY are constructed.
These are clearly also representations of VSUSY, but as such they are in general not irreducible.
However, there is a main difference between what has been done there and the discussion in this paper.
While, as previously mentioned, all multiplets of \cite{Kato:2005gb} and \cite{Kato:2008dw} are constructed in Euclidean space, which is the natural setting in the context of the twisted topological theories,
in this paper we are interested in representations of VSUSY in Minkowski space.
Therefore, in this section we will first exhibit an on-shell VSUSY multiplet with $Y=1/2$, which will be shown in Appendix C to arise from a truncation of the Euclidean multiplet given in \cite{Kato:2005gb}. Then we will discuss whether this multiplet could be consistently rotated to Minkowski space. Finally, we will rederive the VSUSY multiplet directly in VSUSY superspace. The fields in the $Y=1/2$ VSUSY multiplet satisfy dynamical equations of motion with mass, which however cannot be derived from a Lagrangian. Comments about this issue from the point of view of the truncation of the TSUSY multiplet are given in Appendix C.
\subsection{On-shell $Y=1/2$ multiplet}
Consider the set of fields $(a, \tilde b_\m, c^-_{\m\n})$ where the first one is an even scalar, the second an odd vector and the third an even selfdual two-tensor in Euclidean or Minkowski space, according to the definitions:
\begin{eqnarray}
&&c^{\pm}_{\m\n}=\mp\frac{1}{2}\e_{\m\n\rho\s} c^{\pm \rho\s} ~~~~{\rm (Euclidean)}\,, \cr
 &&c^{\pm}_{\m\n}=\mp\frac{i}{2}\e_{\m\n\rho\s} c^{\pm \rho\s} ~~~~{\rm (Minkowski)}\,,
\label{selfdualityI}
\end{eqnarray}
or equivalently
\begin{equation}
 c^{\pm}_{\m\n}=\frac{1}{4} \delta^{\pm}_{\m\n,\rho\s} c^{\pm \rho\s}\,,
\label{selfdualityII}
\end{equation}
where
\begin{eqnarray}
&&\delta^{\pm}_{\m\n,\rho\s} =\delta_{\m\rho}\delta_{\n\s}-\delta_{\m\s}\delta_{\n\rho}\mp  \e_{\m\n\rho\s}~~~~{\rm (Euclidean)}\,,\cr
&&\delta^{\pm}_{\m\n,\rho\s} =\eta_{\m\rho}\eta_{\n\s}-\eta_{\m\s}\eta_{\n\rho}\mp i \e_{\m\n\rho\s}~~~~{\rm (Minkowski)}
\label{projectors}
\end{eqnarray}
is, up a constant, a projector on the (anti)selfdual space.
Therefore, the main difference between the Minkowski and the Euclidean case is that, with Minkowskian signature, (anti)selfdual fields are necessarily complex and the complex conjugate of a selfdual field is antiselfdual. As a result, a model featuring only selfdual or only antiselfdual fields in Minkowski space cannot exist.
Forgetting for a moment about this issue, we consider the VSUSY multiplet in Table \ref{tablevectormultipletI}.
\begin{table}[h!]
\begin{center}
\begin{tabular}{|c|c|c|}
    \hline
      & $Q_{5}$ & $Q_{\mu}$\\
     \hline
      & & \\
     $a$ & $\frac{i}{\hat z}\partial^{\sigma}\tilde b_{\sigma}$ & $-\tilde b_{\mu}$ \\
     & &\\
     $\tilde b_{\nu}$ & $i\left(\frac{2}{\hat z}\partial^{\rho}c^{-}_{\nu\rho}-\frac{1}{2}\partial_{\nu}a\right)$ & $2c_{\mu\nu}^{-}+
       \frac{\hat z}{2}\eta_{\mu\nu}a$ \\
       & &\\
     $c_{\nu\rho}^{-}$ & $\frac{i}{4}\delta_{\nu\rho,\sigma\tau}^{-}\partial^{\sigma}\tilde b^{\tau}$ & $-\frac{\hat z}{4}\delta_{\nu\rho,\mu\sigma}^{-}\tilde b^{\sigma}$ \\
      & & \\
    \hline
    \end{tabular}
\end{center}
\caption{Action of the VSUSY charges on the fields (on-shell multiplet with scalar, vector and selfdual two-form).\label{tablevectormultipletI}}
\end{table}

\noindent It is important to observe that this multiplet closes under VSUSY, in both Euclidean and Minkowski signature, when the constraint $P^2=-Z^2$ is imposed by hand, which amounts to imposing on every component field that $\Box=\hat z^2$. Moreover, the odd Casimir is present and has value zero.
The closure of the algebra on the field $c_{\nu\rho}^{-}$ is nontrivial. To check it one has to use the selfduality of the field (\ref{selfdualityI}) and the Schouten identity $\e_{[\m\n\rho\s} \eta_{\tau]\beta}=0$.\\
It is interesting to note that, if one replaces the selfdual field $c^-_{\m\n}$ with an antiselfdual field $c^+_{\m\n}$, the multiplet closes exactly in the same way, i.e. with no difference in signs or factors.\\
The set of dynamical equations with mass
\begin{eqnarray}
&&\Box a=\hat z^2 a\,;\cr
&&\Box \tilde b_\m=\hat z^2 \tilde b_\m\,; \cr
&&\Box c^-_{\m\n}=\hat z^2 c^-_{\m\n}
\end{eqnarray}
is consistent with VSUSY but cannot be naturally derived from an action.\\
Finally, we would like to discuss the possibility of rotating this multiplet to Minkowski space.
Since, as mentioned before, selfdual fields in Minkowski space are necessarily complex and a model featuring only selfdual fields cannot exist, one expects the whole multiplet to be complex in Minkowski space.
This means that there will be a doubling of the number of degrees of freedom and that in fact two multiplets will be present, $(a, \tilde b_\m, c^-_{\m\n})$ and $(a^*,\tilde b_\m^*, c^+_{\m\n} )$ where of course $c^+_{\m\n}=(c^-_{\m\n})^*$. These two multiplets do not mix in the VSUSY transformations and are only related by complex conjugation. This situation is analogous to what happens with the chiral multiplet of ordinary supersymmetry in four dimensions written in complex notation. One can check that, by taking the complex conjugate of the multiplet of Table \ref{tablevectormultipletI}, given in Table \ref{tablevectormultipletII}, and by taking into account the conventions for the transformations of the fields (\ref{fulltrans}) and (\ref{fulltransZ}), one indeed obtains a copy of the VSUSY algebra for the complex conjugate fields, with the only change in the selfduality property of the tensor field, i.e. everything is consistent.
\begin{table}[h]
 \begin{center}
\begin{tabular}{|c|c|c|}
    \hline
       & $\delta_{Q_5}$ & $\delta_{Q_{\mu}}$\\
     \hline
       & &\\
    $a^*$ & $\frac{1}{\hat z}\e \partial^\s \tilde b_\s^*$ & $-i\e^\m \tilde b_\m^*$ \\
    $\tilde b_\n^*$ & $\e(-\frac{2}{\hat z}\partial^\rho c_{\n\rho}^{+} + \frac{1}{2}\partial_\n a^*)$ & $-\e^\m(2i c_{\m\n}^{+} + i \frac{\hat z}{2}\eta_{\m\n} a^*)$\\
    $c^+_{\n\rho}$ & $\frac{1}{4}\e \delta^+_{\n\rho,\s\tau}\partial^\s\tilde b^{*\tau}$ & $-i \frac{\hat z}{4}\e^\m \delta^+_{\n\rho,\m\s}\tilde b^{*\s}$\\
      & &\\
    \hline
    \end{tabular}
\end{center}
\caption{VSUSY transformations of the complex conjugate fields (scalar, vector, antiselfdual two-form). \label{tablevectormultipletII}}
\end{table}

\subsection{$Y=1/2$ multiplet from superspace}
In this section we derive the $Y=\frac{1}{2}$ multiplet of Table \ref{tablevectormultipletI} by superspace techniques using the superspace setup of Section 3.3.
We have seen that this multiplet closes exactly in the same way under vector supersymmetry in Euclidean and Minkowski spacetime. In Minkowski the fields must be complex and the complex conjugate fields transform into each other according to Table \ref{tablevectormultipletII}. However, since VSUSY does not mix the fields and their complex conjugates in Minkowski space, in this section we do not have to worry about this doubling and simply consider the multiplet of Table \ref{tablevectormultipletI} in either Euclidean or Minkowski spacetime.

\subsubsection{Scalar Superfield}
Let us consider a generic scalar superfield:
\begin{eqnarray}
&T(x^{\mu},\theta^{\mu},\theta_{5},z)=a(x,z)+
\tilde{a}(x,z)\theta_{5}+
\tilde b_{\mu}(x,z)\theta^{\mu}+
b_{\mu}(x,z)\theta^{\mu}\theta_{5}+\nonumber\\
&+c_{\mu\nu}(x,z)\theta^{\mu}\theta^{\nu}+
\tilde{c}_{\mu\nu}(x,z)
\theta^{\mu}\theta^{\nu}\theta_{5}
+\tilde d_{\sigma}(x,z)
\epsilon^{\sigma}_{\,\,\,\mu\nu\rho}\,\theta^{\mu}\theta^{\nu} \theta^{\rho}
+d_{\sigma}(x,z)
\epsilon^{\sigma}_{\,\,\,\mu\nu\rho}\, \theta^{\mu}\theta^{\nu} \theta^{\rho}\theta_{5}+\nonumber\\
&+f(x,z) \epsilon_{\mu\nu\rho\sigma}\, \theta^{\mu}\theta^{\nu} \theta^{\rho} \theta^{\sigma}+
\tilde{f}(x,z) \epsilon_{\mu\nu\rho\sigma}\, \theta^{\mu}\theta^{\nu} \theta^{\rho} \theta^{\sigma}\theta_{5},\cr
&
\label{scalarsuperfield}
\end{eqnarray}
where we assume as before that all components can be Taylor-expanded in $z$ as in (\ref{zetaexp}) and
where $c^{(n)}_{\mu\nu}$ and $\tilde c^{(n)}_{\mu\nu}$ are two-forms. It is useful to split them in their self-dual and anti-self-dual parts
\begin{equation}
c^{(n)}_{\mu\nu}=c^{(n)+}_{\mu\nu}+c^{(n)-}_{\mu\nu}\,, \quad \quad \quad \quad \tilde c^{(n)}_{\mu\nu}=\tilde c^{(n)+}_{\mu\nu}+\tilde c^{(n)-}_{\mu\nu},
\label{split}
\end{equation}
according to equations (\ref{selfdualityI}) (\ref{selfdualityII}) and (\ref{projectors}).
\subsubsection{Supercovariant constraints}
The idea is now to impose a covariant constraint on $T$ that will let us work with only the self-dual parts of $c^{(n)}_{\mu\nu}$ and $\tilde c^{(n)}_{\mu\nu}$. We choose the following constraint:
\begin{equation}
\epsilon^{\mu\nu\rho\sigma}D_{\mu}D_{\nu}D_{\rho}D_{\sigma}T=-6i\, \Box T.
\label{c1}
\end{equation}
After a quite long calculation one can show that the following relations must hold for any integer $n=0,1,2...$:
\begin{eqnarray}
c^{(n)+}_{\mu\nu}=0 &\quad\quad\quad\quad &\tilde c^{(n)+}_{\mu\nu}=\frac{i}{8}\delta_{\mu\nu,\rho\sigma}^{+}\partial^{\rho}\tilde b^{\sigma (n)}
\nonumber\\
\tilde d^{(n)}_{\mu}=\frac{1}{12}\tilde b^{(n+1)}_{\mu} &\quad\quad\quad\quad&         d^{(n)}_{\mu}=\frac{1}{12}b_{\mu}^{(n+1)}+\frac{i}{24}\partial_{\mu}a^{(n+1)}+\frac{1}{6}\partial^{\nu}c_{\mu\nu}^{(n)-}\nonumber\\
f^{(n)}=\frac{i}{96}\,\Box a^{(n)}&\quad\quad\quad\quad &\tilde f^{(n)}=\frac{i}{96}\,\left(\Box\tilde a^{(n)}+ \partial^{\sigma}\tilde b^{(n+1)}_{\sigma}\right)\, ,
\label{expscalar}
\end{eqnarray}
We are left with six infinite sets of independent complex component fields whose VSUSY transformations are given in Table \ref{tableinfinite}.
\begin{table}[h]
\begin{center}
\begin{tabular}{|c|c|c|c|}
    \hline
      & $Q_{5}$ & $Q_{\mu}$ & $Z$\\
     \hline
      & & &\\
    $a^{(n)}$ & $-\tilde a^{(n)}$ & $-\tilde b_{\mu}^{(n)}$ & $i\,a^{(n+1)}$\\
    $\tilde a^{(n)}$ & $-\frac{i}{2}a^{(n+1)}$ & $b_{\mu}^{(n)}-\frac{i}{2}\partial_{\mu}a^{(n)}$ & $i\,\tilde a^{(n+1)}$ \\
    $\tilde b_{\nu}^{(n)}$ & $-b_{\nu}^{(n)}-\frac{i}{2}\partial_{\nu}a^{(n)}$ & $2c_{\mu\nu}^{(n)-}+\frac{i}{2}\eta_{\mu\nu}a^{(n+1)}$ &$i\, \tilde b_{\nu}^{(n+1)}$\\
    $b_{\nu}^{(n)}$ & $\frac{i}{2}\partial_{\nu}\tilde a^{(n)}-\frac{i}{2}\tilde b_{\nu}^{(n+1)}$ & $-\frac{i}{4}\delta^{+}_{\mu\nu,\rho\sigma}\partial^{\rho}\tilde b^{\sigma (n)}-\frac{i}{2}\partial_{\mu}\tilde b_{\nu}^{(n)}-2\tilde c_{\mu\nu}^{(n)-}-\frac{i}{2}\eta_{\mu\nu}\tilde a^{(n+1)}$ & $i\,b_{\nu}^{(n+1)}$ \\
    $c_{\nu\rho}^{(n)-}$ & $\frac{i}{8}\delta^{-}_{\nu\rho,\sigma\tau}\partial^{\sigma}\tilde b^{\tau (n)}-{\tilde c}_{\nu\rho}^{(n)-}$ & $-\frac{i}{4}\delta^{-}_{\nu\rho,\mu\sigma}\tilde b^{\sigma(n+1)}$ & $i\,c_{\nu\rho}^{(n+1)-}$\\
    ${\tilde c}_{\nu\rho}^{(n)-}$ & $-\frac{i}{8}\delta^{-}_{\nu\rho,\sigma\tau}\partial^{\sigma} b^{\tau (n)}-\frac{i}{2}c_{\nu\rho}^{(n+1)-}$ & $\frac{i}{4}\delta^{-\,\,\,\,\,\,\,\sigma}_{\nu\rho,\mu}(\frac{i}{4}\partial_{\sigma}a^{(n+1)}+b_{\sigma}^{(n+1)}+\partial^{\tau}c_{\sigma\tau}^{(n)-})
    -\frac{i}{2}\partial_{\mu}c_{\nu\rho}^{(n)-}$ & $i\,{\tilde c}_{\nu\rho}^{(n+1)-}$ \\
     & & &\\
    \hline
    \end{tabular}
\end{center}
\caption{Action of VSUSY charges on the six infinite sets of component fields solving constraint (\ref{c1}) (two sets of scalars, two of vectors and two of selfdual two-forms).\label{tableinfinite}}
\end{table}

\noindent Now we impose another covariant constraint involving $D_{5}$ in order to further reduce our multiplet:
\begin{equation}
Z D_{5}T=-i\partial^{\mu}D_{\mu}T\, .
\label{c2}
\end{equation}
It is not hard to prove that this constraint implies that $\Box T=Z^{2}T$, so that the higher order components of the fields are related to the lower ones via relations similar to those appearing in (\ref{higher}) and the odd Casimir is present and has value zero. In practice, we need to consider only the components of order 0 and 1 in the $z$ expansion. One can show that the constraint (\ref{c2}) yields:
\begin{eqnarray}
&\tilde a^{(1)}=-\partial^{\mu}\tilde b_{\mu}^{(0)}\,; \quad\quad\quad& \Box \tilde a^{(0)}=-\partial^{\mu}\tilde b_{\mu}^{(1)}\,;\nonumber\\
&b_{\mu}^{(1)}=-2\partial^{\nu}c_{\mu\nu}^{(0)-}\,; \quad\quad\quad& \Box b_{\mu}^{(0)}=-2\partial^{\nu}c_{\mu\nu}^{(1)-}\,; \nonumber\\
&\tilde c_{\mu\nu}^{(0)-}=-\frac{i}{8}\delta_{\mu\nu,\rho\sigma}^{-} \partial^{\rho}\tilde b^{\sigma (0)}\,;\quad\quad\quad&
\tilde c_{\mu\nu}^{(1)-}=-\frac{i}{8}\delta_{\mu\nu,\rho\sigma}^{-} \partial^{\rho}\tilde b^{\sigma (1)}\,.
\end{eqnarray}
Now, if we want to consider $Z$ not as an operator but as a number, indicated as usual with $\hat z$, it is sufficient to impose that $ZT=\hat z T$.
We have then reduced our multiplet to three independent on-shell component fields, the scalar $a^{(0)}$, the vector $\tilde b_{\mu}^{(0)}$ and the selfdual two-form
$c_{\mu\nu}^{(0)-}$. Their VSUSY transformations coincide with the ones in Table \ref{tablevectormultipletI} with the identification $a^{(0)}\equiv a$ and similar for the other fields.

\section{Deriving actions from superspace}
In this section we initiate the study of VSUSY-invariant actions from the superspace point of view.
As a simplification, we choose as before $\tilde Z=-Z$.
Moreover, we will always consider $Z$ as a number, denoted by $\hat z$ to distinguish it from the bosonic coordinate $z$. Therefore, we will not derive the action for the four-spinor multiplet with nontrivial central charge operator discussed in Section 3.2. To derive that, the method of Sohnius \cite{Sohnius:1978fw} should be implemented and we choose to leave this issue for future work. 

Let us first set the dimensions (in mass units) of the coordinates and the covariant derivatives:
\begin{eqnarray}
&[x^{\mu}]=-1\,;\quad[\theta^{\mu}]=-\frac{1}{2}\,;\quad [\hat z]=1\,;\nonumber\\
&[dx^{\mu}]=-1\,;\quad [d\theta]=\frac{1}{2}\,;\quad[\partial_{\mu}]=1\,;\quad [D_{\mu}]=\frac{1}{2}\,; \quad [D_{5}]=\frac{1}{2}\,.
\end{eqnarray}
We define a general action in the superspace as
\begin{equation}
S=\int \,d^{4}x\,d^{4}\theta\,d\theta_{5}\,\mathcal{L}\,.
\label{action}
\end{equation}
Since $S$ has to be even and dimensionless, we have that $\mathcal{L}$ must be odd and of dimension $3/2$.
We choose the dimension of the spinorial superfields $\Phi$, defined in (\ref{spinorsuperfield}), and $\tilde \Lambda$, related to $\Phi$ by (\ref{first}), as
\begin{equation}
[\Phi]=\frac{3}{2}\,, \quad \quad \quad \quad [\tilde\Lambda]=2\,,
\end{equation}
from which one can derive the dimensions of their component fields
\begin{equation}
[\phi^{(0)}]=\frac{3}{2}\,;\quad[\tilde\phi^{(0)}]=2\,;\quad [\tilde\lambda^{(0)}]=2\quad[\lambda^{(0)}]=\frac{5}{2}\,.
\end{equation}
In the same way, we choose a dimension for the scalar superfield $T$ defined in (\ref{scalarsuperfield})
\begin{equation}
[T]=1 ,
\end{equation}
from which one can derive the dimensions of its component fields
\begin{equation}
[a^{(0)}]=1\,,\quad[\tilde{a}^{(0)}]=\frac{3}{2}\,,\quad [\tilde{b}_{\mu}^{(0)}]=\frac{3}{2}\,,\quad[b_{\mu}^{(0)}]=2\,,\quad[c_{\mu\nu}^{(0)-}]=2\,,\quad[\tilde{c}_{\mu\nu}^{(0)-}]=\frac{5}{2}\,.
\end{equation}
Now, since our goal is to build a VSUSY invariant Lagrangian $\mathcal{L}$, we need to set the conventions for the Dirac conjugates
of the superfields. As much as we have chosen, in Section 2.1, our supersymmetry parameters $\e$ and $\e_\m$ to be real by convention, we now choose the superspace coordinates to be real
\begin{equation}
(\theta_{\mu})^*=\theta_{\mu}\,,\quad \quad \quad \quad (\theta_{5})^*=\theta_{5}\,.
\end{equation}
With the conventions adopted in (\ref{conv}) and (\ref{fermionbil}), one can see that the expansion of the bar of the spinor superfield $\Phi$ becomes
\begin{eqnarray}
&\bar\Phi\equiv i\Phi^{\dagger}\gamma_{0}=\bar{\phi}^{(0)}-\bar{\tilde\phi}^{(0)}\theta_{5}-\bar{\tilde\psi}^{(0)}_{\mu}\theta^{\mu}-
{\bar\psi}^{(0)}_{\mu}\theta^{\mu}\theta_{5}
-\bar{\xi}^{(0)}_{\mu\nu}\theta^{\mu}\theta^{\nu} + \bar{\tilde\xi}^{(0)}_{\mu\nu}\theta^{\mu}\theta^{\nu}\theta_{5}+\nonumber\\
&+\bar{\tilde\chi}_{\mu\nu\rho}^{(0)}\theta^{\mu}\theta^{\nu}\theta^{\rho}+\bar{\chi}_{\mu\nu\rho}^{(0)}\theta^{\mu}\theta^{\nu}\theta^{\rho}\theta_{5}
+\bar{\zeta}_{\mu\nu\rho\sigma}^{(0)}\theta^{\mu}\theta^{\nu}\theta^{\rho}\theta^{\sigma}-
\bar{\tilde\zeta}_{\mu\nu\rho\sigma}^{(0)}\theta^{\mu}\theta^{\nu}\theta^{\rho}\theta^{\sigma}\theta_{5}\,,
\end{eqnarray}
where (cfr. (\ref{exp}) and (\ref{sup}))
\begin{eqnarray}
\bar{\tilde\psi}^{(0)}_{\mu}=-\bar{\tilde\lambda}^{(0)}\gamma_{\mu}&\quad\quad\quad\quad&{\bar\psi}^{(0)}_{\mu}=
-\bar{\lambda}^{(0)}\gamma_{\mu}+\frac{i}{2}\partial_{\mu}\bar{\phi}^{(0)}\nonumber\\
\bar{\xi}^{(0)}_{\mu\nu}=\frac{\hat z}{4}\bar{\phi}^{(0)}\gamma_{\mu\nu}&\quad\quad\quad\quad&
\bar{\tilde\xi}^{(0)}_{\mu\nu}=\frac{\hat z}{4}\bar{\tilde\phi}^{(0)}\gamma_{\mu\nu} +\frac{i}{2}\partial_{[\mu}\bar{\tilde\lambda}^{(0)}\gamma_{\nu]}\nonumber\\
\bar{\tilde\chi}_{\mu\nu\rho}^{(0)}=-\frac{\hat z}{12} \bar{\tilde\lambda}^{(0)}\gamma_{\mu\nu\rho}&\quad\quad\quad\quad&
\bar{\chi}_{\mu\nu\rho}^{(0)}=-\frac{\hat z}{12} \bar{\lambda}^{(0)}\gamma_{\mu\nu\rho}+\frac{\hat z}{8} i\partial_{[\mu}\bar{\phi}^{(0)}\gamma_{\nu\rho]}\nonumber\\
\bar{\zeta}_{\mu\nu\rho}^{(0)}=\frac{\hat{z}^{2}}{96}\bar{\phi}^{(0)}\gamma_{\mu\nu\rho\sigma}&\quad\quad\quad\quad&
\bar{\tilde\zeta}_{\mu\nu\rho}^{(0)}=\frac{\hat{z}^{2}}{96}\bar{\tilde\phi}^{(0)}\gamma_{\mu\nu\rho\sigma}
+\frac{\hat z}{24}i\partial_{[\mu}\bar{\tilde\lambda}^{(0)}\gamma_{\nu\rho\sigma]}\,.
\end{eqnarray}
Analogously, for the complex conjugate of the scalar superfield $T$ we have
\begin{eqnarray}
&\bar T=\bar{a}^{(0)}-\bar{\tilde a}^{(0)}\theta_{5}-\bar{\tilde b}^{(0)}_{\mu}\theta^{\mu}-
{\bar b}^{(0)}_{\mu}\theta^{\mu}\theta_{5}
-\bar{c}^{(0)}_{\mu\nu}\theta^{\mu}\theta^{\nu} + \bar{\tilde c}^{(0)}_{\mu\nu}\theta^{\mu}\theta^{\nu}\theta_{5}+
\bar{\tilde d}_{\sigma}^{(0)}\epsilon^{\sigma}_{\,\,\mu\nu\rho}\theta^{\mu}\theta^{\nu}\theta^{\rho}\nonumber\\
&+\bar{d}_{\sigma}^{(0)}\epsilon^{\sigma}_{\,\,\mu\nu\rho}\theta^{\mu}\theta^{\nu}\theta^{\rho}\theta_{5}
+\bar{f}^{(0)}\epsilon_{\mu\nu\rho\sigma}\theta^{\mu}\theta^{\nu}\theta^{\rho}\theta^{\sigma}-
\bar{\tilde f}^{(0)}\epsilon_{\mu\nu\rho\sigma}\theta^{\mu}\theta^{\nu}\theta^{\rho}\theta^{\sigma}\theta_{5}\,,
\end{eqnarray}
where (cfr. (\ref{expscalar}))
\begin{eqnarray}
\bar{c}^{(0)-}_{\mu\nu}=0 &\quad\quad\quad\quad &\bar{\tilde c}^{(0)-}_{\mu\nu}=-\frac{i}{8}\delta_{\mu\nu,\rho\sigma}^{-}\partial^{\rho}
\bar{\tilde b}^{\sigma (0)}\nonumber\\
\bar{\tilde d}^{(0)}_{\mu}=\frac{i}{12}\hat z\bar{\tilde b}^{(0)}_{\mu} &\quad\quad\quad\quad& \bar{d}^{(0)}_{\mu}=\frac{i}{12} \hat z \bar{b}_{\mu}^{(0)}+\frac{1}{24}\hat z \partial_{\mu}\bar{a}^{(0)}+\frac{1}{6}\partial^{\nu}\bar{c}_{\mu\nu}^{(0)+}\nonumber\\
\bar{f}^{(0)}=-\frac{i}{96}\,\Box \bar{a}^{(0)}&\quad\quad\quad\quad &\bar{\tilde f}^{(0)}=-\frac{i}{96}\,\left(\Box\bar{\tilde a}^{(0)}+ i\hat z\partial^{\sigma}\bar{\tilde b}^{(0)}_{\sigma}\right)\, .
\end{eqnarray}
\subsection{Klein-Gordon-type action}
Here we work with the superfields $\Phi$ and $\tilde\Lambda$ defined in Section 4 and satisfying only constraint (\ref{first}). Since $Z$ is treated as a number, we are actually working with four component fields, $\phi^{(0)}$, $\tilde\phi^{(0)}$, $\tilde\lambda^{(0)}$ and $\lambda^{(0)}$, transforming as shown in Table \ref{tablenine}.\\
From (\ref{action}) it is clear that the only term of the Lagrangian that survives integration is the one of highest order in $\theta_{\mu}$ and $\theta_{5}$, denoted by ${\mathcal L}|_{\Theta^{4}\theta_{5}}$:
\begin{equation}
S=\int \,d^{4}x\,d^{4}\theta\,d\theta_{5}\,{\mathcal L}|_{\Theta^{4}\theta_{5}} \,\Theta^{4}\theta_{5}=\int \,d^{4}x\,{\mathcal L}|_{\Theta^{4}\theta_{5}}\,,
\end{equation}
where we define
\begin{equation}
\Theta^{4}\equiv\frac{1}{4!}\e_{\mu\nu\rho\sigma}\theta^{\mu}\theta^{\nu}\theta^{\rho}\theta^{\sigma}.
\end{equation}
We start by building the simplest combinations of $\Phi$ and $\tilde\Lambda$ that give odd Lagrangian terms of dimension $3/2$
\begin{equation}
{\mathcal L}_{mass}=\frac{i}{\hat{z}^{2}}\bar\Phi\tilde\Lambda\,, \quad \quad \quad \quad {\mathcal L}_{kin}=\frac{1}{\hat{z}^{3}}\bar\Phi\dslash D_{5}\Phi\,.
\label{lagr1}
\end{equation}
After some lengthy calculations one can show that these two terms give the following contributions
\begin{eqnarray}
&{\mathcal L}_{mass}|_{\Theta^{4}\theta_{5}}=2\left(\bar{\tilde\phi}^{(0)}\gamma_{5}\tilde\lambda^{(0)} - \bar{\phi}^{(0)}\gamma_{5}\lambda^{(0)}  +
\bar{\lambda}^{(0)}\gamma_{5}\phi^{(0)} - \bar{\tilde\lambda}^{(0)}\gamma_{5}\tilde\phi^{(0)}\right)\,,\\
&{\mathcal L}_{kin}|_{\Theta^{4}\theta_{5}}=-\frac{2}{{\hat z}^{2}}\left(\bar{\tilde\phi}^{(0)}\gamma_{5}\,\Box\,\tilde\lambda^{(0)} - \bar{\phi}^{(0)}\gamma_{5}\,\Box\,\lambda^{(0)}  + \bar{\lambda}^{(0)}\gamma_{5}\,\Box\,\phi^{(0)} - \bar{\tilde\lambda}^{(0)}\gamma_{5}\,\Box\,\tilde\phi^{(0)}\right)\,.
\end{eqnarray}
Now we take the following Lagrangian
\begin{equation}
{\mathcal L_{1}}\equiv {\mathcal L}_{kin}+{\mathcal L}_{mass}\,.
\end{equation}
Integrating it over the supercoordinates and removing the constant factor $\frac{2}{\hat z^2}$, we obtain the action
\begin{equation}
S_{1}=\int d^{4}x~ (\bar{\phi}^{(0)}\gamma_{5}(\Box-\hat{z}^{2})\lambda^{(0)} - \bar{\tilde\phi}^{(0)}\gamma_{5}(\Box-\hat{z}^{2})\tilde\lambda^{(0)}  + \bar{\tilde\lambda}^{(0)}\gamma_{5}(\Box-\hat{z}^{2})\tilde\phi^{(0)} - \bar{\lambda}^{(0)}\gamma_{5}(\Box-\hat{z}^{2})\phi^{(0)}),
\label{actiontable}
\end{equation}
which gives rise, as expected, to Klein-Gordon equations of motion for the component fields
with the usual identification $\hat z=m$.\\
To avoid confusion, some comments are in order. First of all we remind the reader that (\ref{actiontable})  is the action for the multiplet given in Table \ref{tablenine}.
This multiplet was for us an intermediate step in deriving the spinning particle multiplet, given in Table \ref{tablespinning}, compatible with a set of Dirac equations. This step will be done at the level of superfields in the next section. However, as the reader might remember, another $Y=0$ multiplet with a Klein-Gordon type action appears in this paper, namely in Section 3.2. We would like to stress that action (\ref{action4}) presented there is different from (\ref{actiontable}), since it has diagonal kinetic terms for the physical fields and off-diagonal ones for the ghost fields, while in (\ref{actiontable}) all terms are off-diagonal. As a result, action (\ref{action4}) is compatible with  Majorana condition for the spinors while (\ref{actiontable}) is not. As already mentioned before, the method of Sohnius \cite{Sohnius:1978fw} for deriving actions from an extended superspace with central charge must be used to derive (\ref{actiontable}) and we leave this for future work.
\subsection{Spinning particle Dirac-type action}
We know that the component fields appearing in the previous subsection do not satisfy Dirac-type equations of motion. Their VSUSY transformations are not compatible with Dirac equations. However, we have also shown that, if we take some linear combinations of those fields (\ref{lincomb}), the resulting multiplet is compatible with the Dirac equations.
The idea is to now understand how these linear combinations can be implemented at the level of superfields. \\
In order to do so, we start by considering a new superfield $\Psi$, built out of $\Phi$ and $\tilde\Lambda$
\begin{equation}
\Psi= -\frac{i}{2}\Phi+\frac{1}{\hat z}D_{5}\tilde\Lambda\,.
\end{equation}
Now, we make a redefinition of the field components inside $\Psi$ according to (\ref{lincomb}). This leads to the following expansion for $\Psi$
\begin{eqnarray}
&\Psi=\psi_{1}+\sqrt{\frac{\hat z}{2}}\tilde\psi_{2}\theta_{5}+\frac{1}{\sqrt{2\hat z}}\left((z\gamma_{\mu}-\partial_{\mu})\gamma_{5}\tilde\psi_{1}
+i\partial_{\mu}\tilde\psi_{2}\right)\theta^{\mu}+\frac{1}{2}(z\gamma_{\mu}-\partial_{\mu})\gamma_{5}\psi_{2}\theta^{\mu}\theta_{5}+\nonumber\\
&-\left(\frac{\hat z}{4}\gamma_{\mu\nu}\psi_{1}\+\frac{1}{2}\gamma_{\mu}\partial_{\nu}\psi_{1}+\frac{i}{2}\gamma_{5}\gamma_{\mu}\partial_{\nu}\psi_{2}\right)
\theta^{\mu}\theta^{\nu} - \left(\frac{\hat z}{4}\sqrt{\frac{\hat z}{2}}\gamma_{\mu\nu}\tilde\psi_{2}+\frac{1}{2}\sqrt{\frac{\hat z}{2}}\gamma_{\mu}\partial_{\nu}\tilde\psi_{2}\right)\theta^{\mu}\theta^{\nu}\theta_{5}+\nonumber\\
&+\sqrt{\frac{\hat z}{2}}\left(\frac{\hat z}{12}\gamma_{5}\gamma_{\mu\nu\rho}\tilde\psi_{1}+\frac{1}{4}\gamma_{5}\gamma_{\mu\nu}\partial_{\rho}\tilde\psi_{1}
-\frac{i}{4}\gamma_{\mu\nu}\partial_{\rho}\tilde\psi_{2} \right)\theta^{\mu}\theta^{\nu}\theta^{\rho}
+\frac{\hat z}{8}\gamma_{5}\left(\frac{\hat z}{3}\gamma_{\mu\nu\rho}\psi_{2} + \gamma_{\mu\nu}\partial_{\rho}\psi_{2}\right)\theta^{\mu}\theta^{\nu}\theta^{\rho}\theta_{5}+\nonumber\\
&+\frac{\hat{z}}{4}\left(\hat{z} i\gamma_{5}\psi_{1}+i\gamma_{5}\dslash\psi_{1}-\dslash\psi_{2}\right)\Theta^{4}+
\frac{\hat{z}}{4}\left(\hat{z}\sqrt{\frac{\hat z}{2}}i\gamma_{5}\tilde\psi_{2}+\sqrt{\frac{\hat z}{2}}i\gamma_{5}\dslash\tilde\psi_{2}\right)\Theta^{4}\theta_{5}\,.
\end{eqnarray}
Notice that the dimension of $\Psi$ is $3/2$ so for the component fields we have
\begin{equation}
[\psi_{1}]=[\tilde\psi_{1}]=[\tilde\psi_{2}]=[\psi_{2}]=\frac{3}{2}\,.
\end{equation}
Inspired by (\ref{lagr1}), we now construct other Lagrangian terms of the form
\begin{equation}
{\mathcal L_{box}}=\frac{1}{\hat {z}^{3}}\bar\Psi\dslash D_{5}\Psi\,, \quad \quad \quad \quad {\mathcal L_{slash}}=\frac{i}{\hat{z}^{2}}\bar\Psi \Dslash\Psi\,.
\label{lagr2}
\end{equation}
After some lengthy calculations one can show that these two terms give the following contributions:
\begin{eqnarray}
&{\mathcal L}_{box}|_{\Theta^{4}\theta_{5}}=\frac{i}{\hat z}\left(\bar{\tilde\psi}_{1}\gamma_{5}\,\Box\,\tilde\psi_{1} - \bar{\psi}_{1}\gamma_{5}\,\Box\,\psi_{1} - \bar{\psi}_{2}\gamma_{5}\,\Box\,\psi_{2} + \bar{\tilde\psi}_{2}\gamma_{5}\,\Box\,\tilde\psi_{2}\right)\nonumber\\
&{\mathcal L}_{slash}|_{\Theta^{4}\theta_{5}}=-\frac{i}{\hat z}\left(\bar{\tilde\psi}_{1}\gamma_{5}\,\Box\,\tilde\psi_{1} - \bar{\psi}_{1}\gamma_{5}\,\Box\,\psi_{1} - \bar{\psi}_{2}\gamma_{5}\,\Box\,\psi_{2} + \bar{\tilde\psi}_{2}\gamma_{5}\,\Box\,\tilde\psi_{2}\right)+\nonumber\\
&+4\left(-\bar{\psi}_{1}(\dslash +\hat z)\psi_{2} + \bar{\tilde\psi}_{1}(\dslash +\hat z)\tilde\psi_{2} + \bar{\tilde\psi}_{2}(\dslash +\hat z)\tilde\psi_{1} - \bar{\psi}_{2}(\dslash + \hat z)\psi_{1}\right).
\end{eqnarray}
Now we take the following Lagrangian:
\begin{equation}
{\mathcal L_{2}}\equiv \frac{1}{4}\left({\mathcal L}_{box}+{\mathcal L}_{slash}\right).
\end{equation}
By integrating it over the supercoordinates, we obtain the Dirac-type action (\ref{actionstandard}),
with the usual identification $\hat z=m$. 

\section{Multiplets with $Z=\tilde Z=0$}
In \cite{Casalbuoni:2008ez}, the Casimir operators in the case of vanishing central charges were not studied and left for future work.
However, it is easy to derive VSUSY representations in the case of vanishing central charge via a superspace approach. It is then worth to discuss them here.
Dropping the $z$ coordinate in superspace, the supercharges (\ref{chargessuper}) become
\begin{equation}
\label{VSUSYsuperspaceRealisation}
Q_\mu=\frac{\partial}{\partial\theta^\mu}-\frac{i}{2}\theta_5\frac{\partial}{\partial x^\mu} \,,~~~~~~~~~~~~~~~~
Q_5=\frac{\partial}{\partial\theta_5}-\frac{i}{2}\theta^\mu\frac{\partial}{\partial x^\mu}
\end{equation}
and the associated fermionic covariant derivatives are
\begin{equation}\label{CovariantDerivatives}
D_\mu=\frac{\partial}{\partial\theta^\mu}+\frac{i}{2}\theta_5\frac{\partial}{\partial x^\mu} \,,~~~~~~~~~~~~~~~~~
D_5=\frac{\partial}{\partial\theta_5}+\frac{i}{2}\theta^\mu\frac{\partial}{\partial x^\mu}\,.
\end{equation}
Field representations can be obtained in a straightforward way by imposing the constraints $D_\mu \Phi=0$ and $D_5 \Phi=0$ on a generic superfield $\Phi$. The obtained representations will be the VSUSY analogue of the chiral and antichiral superfields for ordinary supersymmetry. We start with the case of a scalar superfield, but it will be then clear that exactly the same procedure works for a superfield with another Lorentz structure.
\subsection{$D_\mu\Phi=0$ Multiplet}
The constraint $D_\mu \Phi=0$ removes all non-scalar component fields and as a result we get the multiplet in components given in Table \ref{VectorMultiplet}.
\begin{table}[h!]\centering
\begin{tabular}{|c|c|c|}\hline
&$Q_5$&$Q_\mu$\\\hline&&\\[-2ex]
$A$&$i\tilde{B}$&$0$\\
$\tilde{B}$&$0$&$-\partial_\mu A$\\[0.5ex]\hline
\end{tabular}
\caption{Transformation rules for the $D_\mu\Phi=0$ multiplet. \label{VectorMultiplet}}
\end{table}
When one starts with a superfield with extra vectorial or a spinorial indices, both component fields inherit the Lorentz index structure from the superfield. The interacting action
\begin{equation} S=\int d^{4}x  \left( \partial_\mu\bar{A}\partial^\mu\tilde{B} +\partial_\mu\bar{\tilde{B}} \partial^\mu A\right) + (\bar A\tilde{B} + \bar{\tilde{B}}A)V(\bar A A)
\end{equation}
is invariant under the transformations of Table \ref{VectorMultiplet}, where $V$ is an arbitrary analytical function and the bar denotes complex conjugation for scalars, vectors etc. and the Dirac conjugation defined in (\ref{daggerconv}) for spinors. Observe that this action has odd Grassmann parity. A discussion and examples of odd actions at the classical level can be found in \cite{Soroka:1995et}, \cite{Soroka:2001jg}. At the level of quantization odd actions seem problematic, so further investigations would be needed in that direction.\\
One might wonder whether it could be possible to construct an even action for the multiplet given in Table \ref{VectorMultiplet}. Since the fields are complex, a diagonal kinetic term for the ghost of the form $\partial_\m \bar{\tilde B} \partial^\m \tilde B$ is not zero. However, the combination of this with a term of the form $\partial_\m \bar{A} \partial^\m A$ cannot be rendered VSUSY invariant.
Therefore, the only possibility to have both a dynamical physical field and a dynamical ghost is to construct mixed, off-diagonal, kinetic terms for them. Due to the general structure of this multiplet, containing only one physical field and one ghost, the mixed, off-diagonal kinetic terms necessarily feature one physical field and one ghost and are therefore odd.

\subsection{$D_5\Phi=0$ Multiplet}\index{$D\Phi=0$ multiplet}
A similar procedure with the scalar covariant derivative can be applied, resulting in the multiplet in Table \ref{ScalarMultiplet}. A similar multiplet is derived by Kato et al.~\cite{Kato:2005fj}, starting from a twisted topological theory.
\begin{table}[h]\centering
\begin{tabular}{|c|c|c|}\hline
&$Q_5$&$Q_\mu$\\\hline&&\\[-2ex]
$A$&$0$&$i\tilde{F}_\mu$\\
$\tilde{F}_\alpha$&$-\partial_\alpha A$&$-iM_{\mu\alpha}$\\
$M_{\alpha\beta}$&$\partial_{[\alpha}\tilde{F}_{\beta]}$&$ -i\epsilon_{\alpha\beta\mu\gamma}\tilde{K}^\gamma$\\
$\tilde{K}_\alpha$&$\frac{1}{2}\epsilon_{\alpha\beta\gamma\delta}\partial^\beta M^{\gamma\delta}$&$-iH\eta_{\alpha\mu}$\\
$H$&$\partial_\alpha\tilde{K}^\alpha$&$0$\\[0.5ex] \hline
\end{tabular}
\caption{Action of the VSUSY generators on the multiplet with vanishing central charge and $D_5\Phi=0$.\label{ScalarMultiplet}}
\end{table}
One can see by inspection that there exists a submultiplet of the multiplet in Table \ref{ScalarMultiplet}, consisting of a vector and a scalar, shown in Table \ref{IrrScalMultiplet}.
\begin{table}[h]
\begin{center}
 \begin{tabular}{|l|c|c|}
\hline
&$Q_5$&$Q_\mu$\\
\hline
$A$&$0$&$i\tilde{F}_\mu$\\
$\tilde{F}_\nu$, with $\partial_{[\mu}\tilde{F}_{\nu]}=0  $ &$-\partial_\nu A$&$0$\\
\hline
\end{tabular}
\caption{Action of the VSUSY charges on the irreducible vector-scalar multiplet with vanishing central charge.\label{IrrScalMultiplet}}
\end{center}
\end{table}
In order to close the algebra on this submultiplet, one has to impose the condition that $\tilde{F}_\nu$ has zero curl, and thus $\partial_\mu\tilde{F}_\nu=\partial_\nu\tilde{F}_\mu$. This implies that this multiplet can be written in terms of the one in Table \ref{VectorMultiplet}.

\section{Conclusions and outlook}
In this paper we have constructed some field representations of vector supersymmetry with nonvanishing central charge, characterized by superspin $Y=0$ and $Y=1/2$. We have discussed their free dynamics in terms of equations of motion and, when possible, of an action. 
Furthermore, we have developed a superspace setup for vector supersymmetry and we have derived our multiplets in this setting. Concerning the construction of actions by superspace techniques, we have only discussed the simplest case of multiplets where the central charge operator acts diagonally on the fields. We leave the other, more involved case for future work.
We have pointed out and worked out in detail the connection between two of our multiplets and some existing results in the literature, in one case in the context of spinning particle models \cite{Casalbuoni:2008iy} and in the other in the context of topological theories with twisted supersymmetry \cite{Kato:2005gb}.
Finally, we have constructed some representations with vanishing central charge by superspace techniques.
For one of those we have succeeded in writing an invariant interacting action, which is however quite bizzarre, first of all because it is odd. The need of odd actions for certain supersymmetric systems had been pointed out in the literature before \cite{Soroka:1995et} \cite{Soroka:2001jg}, but their quantization remains to our knowledge an open problem.
Another unsettling issue concerning one of our multiplets, with central charge this time, is that, despite the fact the fields are spinors, the equations of motion have to be of Klein-Gordon type to be compatible with vector supersymmetry.
Therefore, by just looking at the action the spacetime group seems to be decoupled from the group with respect to which the fields are spinors and the symmetry group seems to be larger, containing two different Lorentz sectors. In fact, the vector supersymmetry transformations mix the two kinds of Lorentz indices, breaking this apparent larger symmetry.

In general, the representations of even superspin contain only fields of half-integer spin, while the representations of odd superspin contain only fields of integer spin. All representations contain an equal number of physical and ghost degrees of freedom and the vector supersymmetry mixes these two sectors.
Therefore, while vector supersymmetry has a very similar algebraic structure compared to ordinary supersymmetry, it has totally different physics and possible applications. It does not unify fields of half-integer spin with fields of integer spin, but physical fields with ghosts instead. Due to this fact, we mainly think of its role as a technical one, for instance in the context of the renormalization of gauge theories, since its presence will likely lead to cancellation of some divergences.
However, ghost fields play a leading role in the quantization of string theory and in that context there is also an interesting interplay between spacetime and worldsheet supersymmetry in the RNS formulation. Therefore, vector supersymmetry could be expected to arise in some string models.
Finally, as already mentioned in the introduction, vector supersymmetry serves as a good comparison to understand what the essential ingredients in supersymmetry are.

The work presented in this paper is of course only the very first step in the direction of constructing and studying interacting theories with underlying VSUSY. The representations we have constructed here are possibly the simplest ones and a more general approach allowing for a classification of VSUSY representations could also be interesting work for the future.

 \label{ss:conclusions}

\section*{Acknowledgements}
We would like to thank M. Caldarelli, J. Gomis, W. Troost and A. Van Proeyen for useful discussions.
L.T. would like to thank the Galileo Galilei Institute for Theoretical Physics for the hospitality offered during the workshop ``New Perspectives in String Theory" and the INFN for partial support during the completion of this work.

\appendix
\section{$Y=0$ multiplet from the spinning particle multiplet}
In \cite{Casalbuoni:2008iy}, rigid VSUSY is used to construct the action of the massive spinning particle with the method of nonlinear realizations. A quantization procedure respecting VSUSY shows that the degrees of freedom of the system, two four-dimensional Dirac spinors $\Psi_1$ and $\Psi_2$, satisfy two decoupled Dirac equation with the same mass. The VSUSY transformations, given in Table \ref{tablegrass}, mix the components of the two spinors.\\
\begin{table}[h]
\begin{center}
\begin{tabular}{|c|c|c|}\hline
 &$Q_5$&$Q_\mu$\\\hline&&\\[-2ex]
$\Psi_1$& $-\sqrt{\frac{\hat z}{2}}\gamma_5 \Psi_2$& $-\frac{1}{\sqrt{2\hat z}}\gamma_5\Big((m\gamma_\m -\partial_\m)\Psi_1 +i\partial_\m \Psi_2\Big)$\\
$\Psi_2$& $-\sqrt{\frac{\hat z}{2}}\gamma_5 \Psi_1$& $-\frac{1}{\sqrt{2\hat z}}\gamma_5\Big(i\partial_\m \Psi_1 + (m\gamma_\m +\partial_\m)\Psi_2\Big)$\\ \hline
\end{tabular}
\caption{Action of the VSUSY charges on the spinning particle multiplet introduced in \cite{Casalbuoni:2008iy}. \label{tablegrass}}
\end{center}
\end{table}

\noindent The VSUSY invariant action for this multiplet is
\begin{equation}
S=\int d^4x \Big(\bar\Psi_2 \gamma_5 (\dslash- m)\Psi_1 + \bar \Psi_1 \gamma_5 (\dslash +m)\Psi_2\Big)\,.
 \label{actiongrass}
\end{equation}
Inspection of the transformations in Table \ref{tablegrass}, together with the fact that $Q_\m$ and $Q_5$ are anticommuting generators in the VSUSY algebra and are therefore expected to be odd, leads to the conclusion that the two spinors $\Psi_1$ and $\Psi_2$ cannot have a definite Grassmann parity and should be further decomposed in an even and an odd part as follows
\begin{eqnarray}
&\Psi_1 = \psi_1+ \tilde\psi_1\,;\cr
&\Psi_2 = \psi_2+ \tilde\psi_2\,,
\label{decomp}
\end{eqnarray}
where as usual $\psi_1$ and $\psi_2$ are even and $\tilde\psi_1$ and $\tilde\psi_2$ are odd.\\
By using (\ref{decomp}) in Table \ref{tablegrass} and then decomposing the transformations in their even and odd parts one obtains the VSUSY transformations of the multiplet previously given in Table \ref{tablespinorsold}.
The action (\ref{actiongrass}) has also no definite Grassmann parity and should be decomposed into its even and odd parts
\begin{equation}
S=S_{\rm even}+ S_{\rm odd}\,,
\end{equation}
where
\begin{equation}
S_{\rm even}=\int d^4x \left[\bar\psi_2 \gamma_5 (\dslash- m)\psi_1 + \bar{\tilde\psi_2} \gamma_5 (\dslash- m)\tilde\psi_1+ \bar \psi_1 \gamma_5 (\dslash +m)\psi_2+ \bar {\tilde\psi_1} \gamma_5 (\dslash +m)\tilde\psi_2\right]
\end{equation}
and
\begin{equation}
S_{\rm odd}=\int d^4x \left[\bar{\tilde\psi_2} \gamma_5 (\dslash- m)\psi_1 + \bar{\psi_2} \gamma_5 (\dslash- m)\tilde\psi_1+ \bar {\tilde \psi_1}\gamma_5 (\dslash +m)\psi_2+ \bar {\psi_1} \gamma_5 (\dslash+ m)\tilde\psi_2\right]\,.
\end{equation}
One can check that both parts are separately invariant under the transformations in Table \ref{tablegrass} and generate all the equations of motion of the multiplet.
It is then natural to choose the even part as the action for the field theoretical model.
One can easily check that $S_{\rm even}$ is just the rescaling (\ref{rescal}) of action (\ref{actionstandard}).
\section{Solution of the superspace constraint equation}
In this Appendix we would like to sketch how to solve constraint (\ref{first}) introduced in Section 3.3 to obtain the general solution (\ref{exp}).\\
Let us expand the spinorial superfield $\tilde\Lambda$ introduced in (\ref{first}) as follows:
\begin{eqnarray}
\tilde\Lambda(x^{\mu},\theta^{\mu},\theta_{5},z)&=&\,\,\,\,\tilde\Delta(x,z)+
\Delta(x,z)\theta_{5}+ \Delta_{\mu}(x,z)\theta^{\mu}+ \tilde\Delta_{\mu}(x,z)\theta^{\mu}\theta_{5}\cr
&&+\tilde\Delta_{\mu\nu}(x,z)\theta^{\mu}\theta^{\nu}+ \Delta_{\mu\nu}(x,z)\theta^{\mu}\theta^{\nu}\theta_{5}
+\Delta_{\mu\nu\rho}(x,z)\theta^{\mu}\theta^{\nu} \theta^{\rho}\cr
&&+\tilde\Delta_{\mu\nu\rho}(x,z) \theta^{\mu}\theta^{\nu}\theta^{\rho}\theta_{5}
+\tilde\Delta_{\mu\nu\rho\sigma}(x,z) \theta^{\mu}\theta^{\nu}\theta^{\rho}\theta^{\sigma}+
\Delta_{\mu\nu\rho\sigma}(x,z) \theta^{\mu}\theta^{\nu}\theta^{\rho}\theta^{\sigma}\theta_{5}\,.\cr
&&
\label{lambdasuperfield}
\end{eqnarray}
Constraint (\ref{first}), expanded order by order in the odd cordinates $\theta^\m$ and $\theta_5$, gives the following set of equations:
\begin{eqnarray}
& -\tilde\psi_\m=\gamma_\m\tilde\Delta\,;\quad
&\psi_\m+\frac{i}{2} \partial_\m \phi=\gamma_\m\Delta\,; \nonumber\\
& 2\xi_{\m\n} -\frac{i}{2}\eta_{\m\n} \partial_z \phi =\gamma_\m\Delta_{\nu}\,;\quad &-2\tilde\xi_{\m\n}+\frac{i}{2}\partial_\m \tilde\psi_\n +\frac{i}{2}\eta_{\m\n}\partial_z \tilde\phi=\gamma_\m \tilde\Delta_{\nu}\,;\nonumber\\
&-3\tilde\chi_{\m\n\rho}+\frac{i}{2}\eta_{\m\n}\partial_z \tilde\psi_\rho=\gamma_\m\tilde\Delta_{\nu\rho}\,;\quad
&3\chi_{\m\n\rho}+\frac{i}{2}\partial_\m\xi_{\n\rho}-\frac{i}{2}\eta_{\m\n}\partial_z \psi_\rho=\gamma_\m\Delta_{\nu\rho}\,;\nonumber\\
&4\zeta_{\m\n\rho\s}-\frac{i}{2} \eta_{\m\n}\partial_z \xi_{\rho\s}=\gamma_\m \Delta_{\nu\rho\sigma}\,;\quad
&-4\tilde\zeta_{\m\n\rho\s}+\frac{i}{2}\partial_\m\tilde\zeta_{\n\rho\s}+\frac{i}{2}\eta_{\m\n}\partial_z\tilde{\rho\s}=\gamma_\m \tilde\Delta_{\nu\rho\sigma}\nonumber\\
&\frac{i}{2}\eta_{\m\n}\partial_z \tilde\chi_{\rho\s\tau}=\gamma_\m\tilde\Delta_{\nu\rho\sigma\tau}\,;\quad
&-\frac{i}{2}\eta_{\m\n}\partial_z \chi_{\rho\s\tau}+\frac{i}{2}\partial_\m \zeta_{\n\rho\s\tau}=\gamma_\m \Delta_{\nu\rho\sigma\tau}\,.
\label{expandedcons}
\end{eqnarray}
Let us start by considering all the terms of order $0$ in $\theta_5$, given in the left column above.
At order $0$ in $\theta^\m$, we have the first equation in the left column in (\ref{expandedcons}), which is solved by requiring $\tilde\psi_\m$ to be of the form
\begin{equation}
\tilde\psi_\m^\a=(\gamma_\m)^\a_{~\b} \tilde\lambda^\b\,,
\label{eqpsi}
 \end{equation}
with $\tilde\lambda^\a(x,z)$ a generic spinor.\\
At order $1$ in $\theta^\m$, we have the second equation in the left column in (\ref{expandedcons}).
To possibly solve this equation, $\xi_{\m\n}$ must contain at least one gamma matrix.
Therefore we take the following general ansatz
\begin{equation}
 \xi_{\m\n}(z)=\gamma_{[\m} \omega_{\n]}\,,
\label{ansatz}
\end{equation}
where $\omega^\a_{\n}$ is a Rarita-Schwinger field to be determined.\\
By plugging in this ansatz we obtain that the following equality must be valid
\begin{equation}
 \gamma_\n \omega_\m = -\frac{i}{2}\eta_{\m\n}\partial_z \phi + \gamma_\m \delta_\n
\end{equation}
for some Rarita-Schwinger field $\delta^\a_\n$.
Clearly, this happens for
\begin{equation}
\omega_\m= -\frac{i}{4}\gamma_\m \partial_z \phi\,.
\end{equation}
Note that a term in $\xi_{\m\n}$ of the form $\gamma_{[\m} \partial_{\n]} \hat\omega(z)$
would be allowed by symmetry considerations but in general does not solve the equation, apart from the case when it reduces to the structure we have already considered before.\\
To summarize, the solution of the constraint at order $1$ in $\theta^\m$ leads to the following relation between the components $\xi_{\m\n}^\a$ and $\phi^\a$:
\begin{equation}
\xi_{\m\n}=-\frac{i}{4}\gamma_{\m\n}\partial_z \phi\,.
 \label{theta1}
\end{equation}
A similar procedure works up to order $3$ in $\theta^\m$.\\
At order $4$, we see from the last equation in the first column in (\ref{expandedcons}) that only one term is present in the LHS. However, by using the gamma matrix identity
\begin{equation}
\gamma_{\mu\nu\rho}=-i\epsilon_{\mu\nu\rho\sigma}\gamma^{\sigma}\gamma_{5}\,.
\label{gamma5stuff}
\end{equation}
and the Schouten identity,
one can rewrite the order 4 term as follows
\begin{equation}
 \eta_{\m\n}\gamma_{\rho\s\delta}~\partial_z^2 \tilde\lambda~\sim~ \gamma_\m \gamma_5\e_{\n\rho\s\delta}~\partial_z^2 \tilde \lambda
\label{correctform}
\end{equation}
It is then clear that the order 4 term directly satisfies the constraint equation.

Let us now consider the terms of order $1$ in $\theta_5$.\\
The difference with respect to the previous case is that the term $\frac{i}{2}\theta_{5}\partial_\mu$ in the covariant derivative $D_\m$ now plays a role.
However, one could in principle decide to work in the VSUSY analogue of the chiral representation for the superspace covariant derivatives given in (\ref{newcovderivII}). In that case the $\theta_5$ term in $D_\m$ would simply not be present.
Therefore, in that basis the solution of the constraint equations coming from the terms of order $1$ in $\theta_5$ is completely analogue to the case of order $0$ in $\theta_5$.
It is then clear that the solution rewritten in the more symmetric representation for the covariant derivatives (\ref{covdevII}) will just contain an extra correction term coming from the corresponding change of basis (\ref{covdertrans}).
Explicitly, the analogues of eqs. (\ref{eqpsi}) and (\ref{theta1}) become
 \begin{eqnarray}
&\psi_\m=\gamma_\m\lambda -\frac{i}{2}\partial_\m\phi\,,\cr
&\tilde\xi_{\m\n}=-\frac{i}{4}\gamma_{\m\n}\partial_z \tilde\phi -\frac{i}{2}\gamma_{[\m}\partial_{\n]} \tilde\lambda\,.
 \end{eqnarray}
 The correction is of course the second term in the RHS, which can be easily computed by hand.
Note that it also fits the general prescription given in (\ref{forms}).\\
The result is the general solution of constraint (\ref{first}) given in (\ref{exp}).

\section{$Y=1/2$ multiplet from the Kato-Miyake TSUSY multiplet}
The multiplet in Table \ref{tablevectormultipletI} is a consistent truncation of the TSUSY multiplet found by Kato and Miyake in Euclidean spacetime in \cite{Kato:2005gb}. The Kato-Miyake multiplet contains three vectors $\tilde F^\m$,$V_\m$ and $K^\m$, one odd and two even, an odd scalar $\tilde A$ and an odd selfdual two-form $\tilde M^-_{\m\n}$. The action of TSUSY on these fields is given in Table \ref{tablekatomiyake}.
\begin{table}[h!]
\begin{tabular}{|c|c|c|c|}
\hline
&$Q_5$&$Q_\mu$&$Q^+_{\mu\nu}$\\\hline
& & & \\
$V_\alpha$&$\tilde{F}_\alpha$&$\delta_{\alpha\mu}\tilde{A}-\tilde{M}^-_{\mu\alpha}$& $-\delta^+_{\m\n,\a\rho}\tilde F^\rho$\\
$\tilde{F}_\alpha$&$\frac{1}{2}K_\alpha$&$\frac{i}{2}(\delta^{-}_{\mu\alpha,\rho\sigma}\partial^\rho V^\sigma +\delta_{\alpha\mu}\partial^\rho V_\rho - 2\partial_\mu V_\alpha) $&$\frac{1}{2}\delta^+_{\mu\nu,\alpha\rho} K^\rho $\\
$\tilde{A}$&$-\frac{i}{2}\partial^\mu V_\mu$&$\frac{1}{2}K_\mu$&$-\frac{i}{2}\delta^+_{\mu\nu,\rho\sigma}\partial^\rho V^\sigma $\\
$\tilde{M}^-_{\alpha\beta}$&$\frac{i}{2}\delta^{-}_{\alpha\beta,\rho\sigma}\partial^\rho V^\sigma $&$-\frac{1}{2}\delta^{-}_{\alpha\beta,\mu\rho}K^\rho$&$-\frac{i}{2}\delta^+_{\mu\nu,\rho\sigma}\delta_{\alpha\beta,\tau\phantom{\sigma}}^{-\phantom{\mu\nu,}\sigma} \partial^\rho V^\tau $\\
$K_\alpha$&$-i(\partial_\alpha\tilde{A}+\partial^\rho\tilde{M}^-_{\alpha\rho})$&$-i(\delta^{-}_{\mu\alpha,\rho\sigma}\partial^\rho\tilde{F}^\sigma +\delta_{\alpha\mu}\partial^\rho \tilde{F}_\rho)$&$i\delta^+_{\mu\nu,\alpha\rho}(\partial^\rho \tilde{A}+\partial^\sigma \tilde{M}^{-\rho}_{~~~\sigma} )$\\
& & & \\
\hline

\end{tabular}
\end{table}
\begin{table}[h!]
\begin{tabular}{|c|c|}
 \hline
& Z \\
\hline
& \\
$V_\alpha$&$K_\a$\\
$\tilde{F}_\alpha$&$-i(\partial_\alpha \tilde{A}+\partial^\rho \tilde{M}^-_{\alpha\rho})$\\
$\tilde{A}$&$-i\partial^\rho\tilde{F}_\rho$\\
$\tilde{M}^-_{\alpha\beta}$&$i\delta^{-}_{\alpha\beta,\rho\sigma}\partial^\rho \tilde{F}^\sigma$\\
$K_\alpha$&$-\partial^2 V_\alpha$\\
& \\
\hline
\end{tabular}
\caption{Action of the TSUSY charges on the Kato-Miyake multiplet \cite{Kato:2005gb}.\label{tablekatomiyake}}
\end{table}
\vskip 12pt
\noindent In this multiplet $P^2=Z^2$ (note the different convention with respect to this paper) is satisfied on all fields, but no field equations have to be used to close the algebra, the multiplet closes off-shell. Moreover, the odd Casimir is present with value zero.\\ 
We can shorten the multiplet by imposing that the operator $Z$ acts as a number $\hat z$ on all fields. As a result we obtains the set of constraints:
\begin{eqnarray}
&& K_\m = \hat z V_\m\,;\cr
&& -i(\partial_\m \tilde A + \partial^\rho \tilde M^-_{\m\rho})= \hat z \tilde F_\m \,;\cr
&& -i \partial^\rho \tilde F_\rho =\hat z \tilde A\,;\cr
&& i \delta^-_{\m\n,\rho\s}\partial^\rho \tilde F^\s= \hat z \tilde M^-_{\m\n}\,;\cr
&& -\Box V_\m = \hat z K_\m \,.
\label{setcons}
\end{eqnarray}
The first two equations above tell us that the vectors $K_\m$ and $\tilde F_\m$ can be expressed in terms of $\tilde A$, $V_\m$ and $\tilde{M}^-_{\m\n}$. The remaining three equations imply that $\tilde A$, $V_\m$ and $\tilde{M}^-_{\m\n}$ must satisfy the following dynamical equations with mass:
\begin{eqnarray}
&&\Box V_\m = -\hat z^2 V^\m\,;\cr
&&\Box \tilde A = - \hat z^2 \tilde A \,;\cr
&&\Box \tilde M^-_{\m\n}= - \hat z^2 \tilde M^-_{\m\n}\,.
\label{eqns}
\end{eqnarray}
By implementing the first two constraints of (\ref{setcons}) in Table \ref{tablekatomiyake}, one obtains the corresponding table for the truncated multiplet $(\tilde A,V_\m, \tilde{M}^-_{\m\n})$.
In order to have a VSUSY sector that is exactly identical (up to signs due to the different conventions) to our VSUSY multiplet given in Table \ref{tablevectormultipletI}, we rescale the fields as follows
\begin{equation}
\tilde a=i\tilde A, ~~~~~~~b_\m=-\frac{i}{2} \hat z V_\m, ~~~~~~~\tilde{c}^-_{\m\n}= -i\frac{\hat z}{4}\tilde M^-_{\m\n}\,
\end{equation}
and we switch the roles of ghosts and physical fields.\\
The result is our VSUSY multiplet of Table \ref{tablevectormultipletI}, with the extra tensor transformations
\begin{table}[h!]
\begin{center}
\begin{tabular}{|c|c|}
    \hline
      & $Q^+_{\mu\n}$\\
     \hline
      & \\
     $a$ & $\frac{i}{\hat z}\delta^+_{\m\n,\rho\s}\partial^\rho \tilde b^\sigma$\\
     & \\
     $\tilde b_{\a}$ &  $-\frac{i}{2}\delta^+_{\m\n,\a\rho}(-\partial^\rho a + \frac{4}{\hat z}\partial^\gamma c^{-\rho}_{~~~\gamma}$)\\
       &\\
     $c_{\a\b}^{-}$ &  $-\frac{i}{4} \delta^{+}_{\m\n,\rho\s}\delta^{-~~~\s}_{\a\b,\tau}\partial^\rho \tilde b^\tau$\\
      &\\
    \hline
    \end{tabular}
\end{center}
\caption{Action of the antiselfdual tensor charge on the truncated Kato-Miyake multiplet}
\end{table}
\vskip 3pt
The dynamics of the Kato-Miyake multiplet is given by the action
\begin{equation}
S=\int d^4 x \left(V^\m \Box V_\m + 4i \tilde F^\m(\partial_\m \tilde A + \partial^\n \tilde M^-_{\m\n}) + K^\m K_\m\right)\,.
\label {KMaction}
\end{equation}
The field $K^\m$ is auxiliary and all fields satisfy equations of motion with zero mass.
Note that the scalar and two-form fields appear in the action with an off-diagonal first order term.
The fields remaining after our truncation satisfy dynamical equations with mass instead, related to the value of the central charge by a BPS-like constraint.
The truncation procedure leads to constraints directly imposing the equations of motions on the fields, so it is not possible to apply it to action (\ref{KMaction}) to obtain an action for the truncated multiplet.

We have checked whether it was possible to build an action for the truncated multiplet leading to the equations of motion (\ref{eqns}), but we find no suitable VSUSY invariant quadratic structure. 
On the other hand, one could think of an off-diagonal first order structure as in (\ref{KMaction}), in this case necessarily odd. However, this would not lead to the correct dynamical equations and in any case direct inspection shows that a combination of this kind of terms cannot be rendered invariant under the odd vectorial VSUSY charge.
Our conclusion is that the $Y=\frac{1}{2}$ multiplet satisfies dynamical equations of motion with mass that cannot be derived by an action.

\newpage


\begin{thebibliography}{10}

\bibitem{Barducci:1976qu}
A.~Barducci, R.~Casalbuoni  and L.~Lusanna, \emph{{Supersymmetries and the
  pseudoclassical relativistic electron}}, Nuovo Cim. {\bf A35} (1976)
\href{http://dx.doi.org/10.1007/BF02730291}{377}

\bibitem{Casalbuoni:2008ez}
  R.~Casalbuoni, F.~Elmetti, J.~Gomis, K.~Kamimura, L.~Tamassia and A.~Van Proeyen,
  \emph{{Vector Supersymmetry: Casimir operators and contraction from $OSp(3,2|2)$}},
  JHEP {\bf 0901} (2009) 035
  {{\tt arXiv:0812.1982 [hep-th]}}.

\bibitem{Casalbuoni:2009en}
  R.~Casalbuoni, F.~Elmetti, J.~Gomis, K.~Kamimura, L.~Tamassia and A.~Van Proeyen,
  \emph{{Vector Supersymmetry from $OSp(3,2|2)$: Casimir Operators}},
  Fortsch.\ Phys.\  {\bf 57} (2009) 521
  {{\tt arXiv:0901.4862 [hep-th]}}.

  \bibitem{Witten:1988ze}
E.~Witten, \emph{{Topological quantum field theory}}, Commun. Math. Phys. {\bf
  117} (1988)
\href{http://dx.doi.org/10.1007/BF01223371}{353}

\bibitem{Alvarez:1994ii}
M.~Alvarez and J.~M.~F. Labastida, \emph{{Topological matter in
  four-dimensions}}, Nucl. Phys. {\bf B437} (1995)
  \href{http://dx.doi.org/10.1016/0550-3213(94)00512-D}{356--390},
\href{http://arxiv.org/abs/hep-th/9404115}{{\tt arXiv:hep-th/9404115}}

\bibitem{Kato:2005fj}
J.~Kato, N.~Kawamoto  and A.~Miyake, \emph{{N = 4 twisted superspace from
  Dirac-Kaehler twist and off- shell SUSY invariant actions in four
  dimensions}}, Nucl. Phys. {\bf B721} (2005)
  \href{http://dx.doi.org/10.1016/j.nuclphysb.2005.05.024}{229--286},
\href{http://arxiv.org/abs/hep-th/0502119}{{\tt arXiv:hep-th/0502119}}

\bibitem{Birmingham:1988bx}
D.~Birmingham, M.~Rakowski  and G.~Thompson, \emph{{Topological field theories,
  Nicolai maps and BRST quantization}}, Phys. Lett. {\bf B214} (1988)
\href{http://dx.doi.org/10.1016/0370-2693(88)91381-0}{381}

\bibitem{Delduc:1989ft}
F.~Delduc, F.~Gieres  and S.~P. Sorella, \emph{{Supersymmetry of the d = 3
  Chern-Simons action in the Landau gauge}}, Phys. Lett. {\bf B225} (1989)
\href{http://dx.doi.org/10.1016/0370-2693(89)90584-4}{367}


\bibitem{Baulieu:2008at}
L.~Baulieu, G.~Bossard  and A.~Martin, \emph{{Twisted superspace}}, Phys. Lett.
  {\bf B663} (2008)
  \href{http://dx.doi.org/10.1016/j.physletb.2008.03.054}{275--280},
\href{http://arxiv.org/abs/0802.1980}{{\tt arXiv:0802.1980 [hep-th]}}

\bibitem{Casalbuoni:2008iy}
R.~Casalbuoni, J.~Gomis, K.~Kamimura  and G.~Longhi, \emph{{Space-time vector
  supersymmetry and massive spinning particle}}, JHEP {\bf 02} (2008)
  \href{http://dx.doi.org/10.1088/1126-6708/2008/02/094}{094},
\href{http://arxiv.org/abs/0801.2702}{{\tt arXiv:0801.2702 [hep-th]}}

\bibitem{Kato:2005gb}
J.~Kato and A.~Miyake,
  \emph{{Topological hypermultiplet on N = 2 twisted superspace in four
  dimensions}},
  Mod.\ Phys.\ Lett.\  A {\bf 21}, 2569 (2006)
  [arXiv:hep-th/0512269].

\bibitem{Kato:2008dw}
  J.~Kato and A.~Miyake,
  \emph{{Vafa-Witten theory on N=2 and N=4 twisted superspace in four dimensions}},
  JHEP {\bf 0903}, 087 (2009)
  [arXiv:0808.2538 [hep-th]].

\bibitem{Sohnius:1978fw}
  M.~F.~Sohnius,
  \emph{{Supersymmetry and central charges}},
  Nucl.\ Phys.\  B {\bf 138} (1978) 109.



\bibitem{Soroka:1995et}
V.~A.~Soroka,
  \emph{{On action with Grassmann-odd Lagrangian}},
  Phys.\ Atom.\ Nucl.\  {\bf 59}, 1270 (1996)
  [Yad.\ Fiz.\  {\bf 59}, 1327 (1996)]
  [arXiv:hep-th/9507030].


\bibitem{Soroka:2001jg}
  D.~V.~Soroka, V.~A.~Soroka and J.~Wess,
  \emph{{Supersymmetric D = 1, N=1 model with Grassmann-odd Lagrangian}},
  Phys.\ Lett.\  B {\bf 512}, 197 (2001).

\bibitem{thesis}
S.~Knapen, \emph{{Field representations of vector supersymmetry}}, master thesis, K.U.Leuven, 2008-2009.

\bibitem{VanProeyen:1999ni}
A.~Van Proeyen,
  \emph{{Tools for supersymmetry}},
  arXiv:hep-th/9910030.

\end{thebibliography}

\providecommand{\href}[2]{#2}\begingroup\raggedright\endgroup

\end{document}